\documentclass[12pt]{article}

\usepackage{graphicx}

\newcommand{\beq} {\begin{equation}}
\newcommand{\eeq} {\end{equation}}
\newcommand{\what}[1]{\widehat #1}

\newcommand{\lh} {l_h}
\newcommand{\jh} {j_h}
\newcommand{\mh} {m_h}
\newcommand{\lp} {l_p}
\newcommand{\jp} {j_p}
\newcommand{\emp} {m_p}

\newcommand{\bsigma}{\mbox{\boldmath $\sigma$}}
\newcommand{\btau}{\mbox{\boldmath $\tau$}}
\newcommand{\bkey}{{\bf k}}
\newcommand{\bpi}{{\bf p}}
\newcommand{\bqu}{{\bf q}}
\newcommand{\br}{{\bf r}}
\newcommand{\half}{\frac{1}{2}}
\newcommand{\e}[1]{ {\rm e}^{#1} }

\begin{document}
\begin{titlepage}
\thispagestyle{empty}

%\vspace*{2cm}

\begin{center}
{\Large \bf Short-range correlations and meson exchange currents in 
            photonucleon emission}

\vspace{1.5cm}
{\large  M. Anguiano$ ^{\,1,2}$, G. Co'$ ^{\,\,1,2}$,
A.M. Lallena$^{\,3}$ and S.R. Mokhtar$^{\,1,4}$} \\ 
\vspace{1.cm}
{$^{1)}$ Dipartimento di Fisica,  Universit\`a di Lecce, \\
I-73100 Lecce, Italy} \\
\vspace{.5cm}  
{$^{2)}$  Istituto Nazionale di Fisica Nucleare  sez. di Lecce, 
\\ I-73100 Lecce, Italy} \\
\vspace{.5cm}
{$^{3)}$ Departamento de F\'{\i}sica
Moderna, Universidad de Granada, \\
E-18071 Granada, Spain} \\
\vspace{.5cm}
{$^{4)}$  Department of Physics, University of Assiut, \\
        Assiut, Egypt} \\
\end{center}
\vskip 1.5 cm 
\begin{abstract}
  One-nucleon emission processes induced by photon absorption are
  studied by considering short-range correlations effects. At
  energies above the giant resonance region the validity of the direct
  knock-out model has been tested by comparison with continuum Random
  Phase Approximation results. Nucleon re-scattering effects
  have been considered by using an optical potential. The role of the
  electromagnetic convection, magnetization and meson exchange
  currents has been investigated as a function of both excitation
  energy and momentum transfer.  The short-range correlation effects
  have been studied by using various
  correlation functions.  We found that the nucleon photo-emission
  cross section is rather sensitive to the presence of short-range
  correlations at large values of nucleon emission angle. In this
  region, however, the effects of meson exchange currents are even
  larger than those produced by short-range correlations.
\end{abstract} 
\vskip 1.cm
PACS number(s): 21.10.Ft, 21.60.-n
\end{titlepage}
\newpage
\section{INTRODUCTION}
The search for signatures of Short-Range Correlations (SRC) is one of
the nuclear structure hot topics of these recent years. SRC are
produced by the strong repulsive core of the bare nucleon-nucleon
interaction.  Medium-heavy nuclei are usually described with effective
theories and interactions where the effects of the repulsive core, and
therefore of the SRC, have been already tamed. The observation of
effects clearly produced by SRC would set limits to the validity of
the effective theories.

Perhaps the best known and most spectacular effect produced by the
SRC is the increase of the nucleon momentum
distribution at high momentum values with respect to the
Independent Particle Model (IPM) predictions 
\cite{ant88}-\cite{fab01}.
Unfortunately, the nucleon momentum distribution is not directly
observable, therefore one has to search for phenomena sensitive to
this quantity.  

The first place to look for these phenomena is the nuclear ground
state, the most investigated state from both experimental
and theoretical point of view. Recently ground state calculations
based on microscopic interactions, therefore considering SRC, 
have been provided also for medium-heavy 
nuclei \cite{pud97}-\cite{fab00}. It has been found that the
SRC slightly modifies the charge distribution in the nuclear center,
confirming the results of early calculations \cite{bro63,co95}. 
However, the
size of this effect is comparable to that produced by the
coupling of low-lying collective phonons to the single particle wave
functions \cite{ang01} and therefore it is difficult to disentangle.
No other ground state observables showing relevant 
sensitivity to the SRC have been found, 
therefore one has to search for processes
involving the excitation of the nuclear system.

The status of the theory in the description of nuclear excited states
is not so well advanced as in the case of the ground state.  In these
last years we have developed a model which considers SRC in the
description of nuclear excited states \cite{co98}-\cite{mok01}.  Our
work is based on the nuclear matter model of Ref. \cite{fan87} used to
study inclusive responses \cite{fab89} and spectral functions
\cite{ben89}. With respect to this model we made an additional
approximation consisting in cutting the cluster expansion at the first
order in the correlation line.  A detailed presentation of our
approach is given in Ref. \cite{co01} where it has been applied to
the description of the quasi-elastic inclusive responses.  Recently,
we extended the model to investigate (e,e'p) reactions \cite{mok01}, and
in the present paper we use it to study one-nucleon emission processes
induced by photon absorption.

The two main approximations of our model are 
the exclusion of the
diagrams containing more than one correlation line, and the neglect
of collective excitations of the nucleus, such as the surface
vibrations.  The first approximation has been tested in Ref.
\cite{ama98} by comparing our nuclear matter quasi-elastic charge
responses with those obtained with a complete Correlated Basis
Function calculation \cite{fan87,fab89}.
The agreement between the two calculations is excellent.  

The second approximation has limited the application of
our model to situations where collective
excitations can be neglected.  In Ref. \cite{mok00} we calculated
discrete excitations which can be rather well described in terms of
single particle excitations. In \cite{co98,co01} and \cite{mok01} we
applied our model in the quasi-elastic kinematic regime,
dominated by one-particle one-hole (1p-1h) excitations \cite{ama01}.

The characteristic kinematic conditions which in electron scattering
ensured that
collective effects were negligible 
(high values of the momentum transfer for a fixed excitation energy), 
cannot be achieved with real photons. For this reason we felt
necessary to study the importance of collective excitation modes in
photo-reactions. We have done this investigation by using
the continuum Random Phase Approximation (RPA) theory,
and we found that collective excitations can be neglected, 
above the giant resonance region. 
As in the electron scattering case, the largest correction to the
naive mean field model arises from
the rescattering of the emitted nucleon
with the rest nucleus. 
This final state interaction can be described rather well
within an optical model framework.

After establishing validity and limitations of the direct knock-out
model, we have investigated the role of electromagnetic currents
operators used in the calculations.  
In the energy region of interest we found that 
it is safer to use the explicit form of the
convection current rather than employing  
the long wave approximation. Furthermore it is necessary
to include the magnetization current.
The contribution of Meson Exchange Currents (MEC) generated by the
exchange of pions has been investigated with the
 method used in Refs. \cite{ama93,ama94} to study
the quasi-elastic (e,e') responses.  We have obtained MEC effects
smaller than those presented in the literature. However they are
relevant and cannot be neglected.

After testing the validity of the model and the role of the
electromagnetic operators,  
we started our investigation of the SRC. 
In our calculations the SRC act only on the one-body (OB)
operators, therefore MEC and correlations interact only through
the interference between the transition amplitudes. We used purely
scalar (Jastrow) correlation functions fixed in
Fermi Hypernetted Chain (FHNC) calculations 
done with semi-realistic \cite{ari96} and 
realistic \cite{fab00} nucleon-nucleon
interactions. 
The consistency between the hole single particle wave functions
and the correlation functions, is provided by the 
the minimization procedure used in the FHNC calculations.

All our calculations have been done for the
$^{16}$O nucleus. This is a light doubly closed shell nucleus 
and it has been widely investigated from the theoretical
point of view. At the same time a large set of 
experimental data is available in the literature. 
These facts allow us to compare our results  for various
kinematic conditions
with those of other theories and with
experimental data. Furthermore there are also
(e,e'p) data we can use to test our model.

The paper is mainly focused on the numerical results. 
Details of the
correlated model and of the MEC treatment are given in Refs.
\cite{co01} and \cite{ama93,ama94} respectively. Here we discuss
briefly the extension of the model to the case of the photonuclear
reactions in Sects. \ref{xsection} and \ref{nucmod}. The
applications of our model are presented in Sect. \ref{applications}
which has been divided in various parts following the logical path
above described: first we present the investigation about collective
modes (Sect. \ref{rpa}), then about the currents (Sect. \ref{mec})
followed by the study of the SRC effects (Sect. \ref{src}) and finally
by the comparison with the data (Sect. \ref{data}). In the last
section we summarize the results and draw our conclusions.

\section{THE CROSS SECTIONS}
\label{xsection}
In this section we briefly recall the expressions used to evaluate the
cross section for the nucleon emission induced by the absorption of a
photon. Detailed derivations of these expressions can be found
in review articles \cite{gia85} and books \cite{bla52,bof96}.
We work in natural units ($\hbar=c=1,e^2=1/137.04$) and employ the
conventions of Bjorken and Drell \cite{bjo64}. 

The basic quantity to
calculate is the transition amplitude
\beq
R_T(q,\omega)\, = \, \sum_{\eta= \pm 1}\,  
|\langle \Psi_f|\, J_\eta(q) \, 
|\Psi_i \rangle |^2 \,\, \delta (E_f - E_i - \omega) \, ,
\label{rt} 
\eeq
where we have indicated with $|\Psi_i \rangle$ and $|\Psi_f \rangle$
the initial and final nuclear states and with $E_i$ and $ E_f$ their
energies. 
In the above expression we show the dependence of 
transition amplitude on $q$ and $\omega$,
the modulus of the momentum and the energy
transferred to the nucleus. 
In electron scattering processes, by neglecting the electron rest
mass, one has 
$\epsilon_i-\epsilon_f \leq q \leq 
\epsilon_i+\epsilon_f$ where $\epsilon_i$ and $\epsilon_f$ are the
initial and final energies of the electron.
In processes involving real photons one has
$\omega=q=\epsilon_i-\epsilon_f=E_f-E_i$. Another difference between
processes involving electrons and real photons is that in the last
case the longitudinal term of the electromagnetic current, related to
the nuclear charge distribution, does not contribute. For this reason
in Eq. (\ref{rt}) only the transverse components of the current
appear
\beq
J_{\pm}=\mp \frac{1}{\sqrt{2}} 
\left(J_x \pm i J_y  \right) \, .
\label{jpm}
\eeq

The nuclear initial state is characterized by the total angular
momentum and parity of the system. Since 
we restrict our calculations to doubly closed shell
nuclei we
immediately consider that the nuclear ground state has zero angular
momentum and positive parity 
$| \Psi_i \rangle \equiv |\Psi;J_i M_i ;\Pi_i \rangle =
|\Psi;00;+1\rangle$. 
In processes where the emitted nucleon is detected,  
the nuclear final state can be described as \cite{mok01,co85,co87}
\begin{eqnarray}
\nonumber
|\Psi_f\rangle &=& \frac{4 \pi}{|{\bf p}|}
\sum_{\lp \mu_p} \sum_{\jp \emp} \sum_{JM,\Pi}
         i^{\lp} Y^*_{\lp \mu_p} (\what{\bpi}) 
         \langle \lp \mu_p \half \sigma | \jp \emp \rangle \\
&~&      \langle \jp \emp \jh \mh| J M \rangle \, 
|\Psi; J M; \Pi; (\lp \jp \emp \epsilon_p,\lh \jh \mh \epsilon_h) 
\rangle \, .
\label{psif}
\end{eqnarray}
In the above equation $|\Psi; J M; \Pi; (\lp \jp \emp \epsilon_p, \lh
\jh \mh \epsilon_h) \rangle$ describes the excited state of the A
nucleons system with total angular momentum $J$, $z-$axis projection
$M$ and parity $\Pi$. This state is composed by a particle in the
continuum wave characterized by orbital and total angular momenta
$\lp$ and $\jp$, respectively, with projection $\emp$, energy
$\epsilon_p$ and momentum ${\bf p}$, and a residual nucleus with hole
quantum numbers $\lh$ $\jh$, $\mh$ and $\epsilon_h$.  With the symbol
$Y_{l \mu}$ we indicated the spherical harmonics and with $\langle l_a
m_a l_b m_b | J M \rangle$ the Clebsch-Gordan coefficients
\cite{edm57}.  Using the above definitions we express the transition
amplitude as \cite{mok01,co85,co87}
\begin{eqnarray}
\nonumber
&~& 
R_T(q,\omega)\,= \frac{32 \pi^3}{|{\bf p}|^2} \sum_{\sigma m_h}
\sum_{J \Pi \lp \jp} \sum_{J' \Pi' \lp' \jp'}
\sum_{\eta \pm 1} (-i)^{\lp-\lp'-J+J'}
\frac{1}{\sqrt{2J+1}} \frac{1}{\sqrt{2J'+1}} 
\\ 
\nonumber
&~& \langle \lp  \mu_p \half \sigma | \jp \emp \rangle 
     \langle \jp \emp \jh \mh| J -\eta \rangle
\langle \lp' \mu_p \half \sigma 
| \jp'  \emp' \rangle 
     \langle \jp' \emp \jh \mh| J' -\eta \rangle 
\\
\nonumber
&~& 
\left[ \frac {(\lp+\mh +\eta+\sigma)! } {(\lp-\mh -\eta-\sigma)! }
\frac {(\lp' +\mh +\eta +\sigma)! } {(\lp' -\mh -\eta-\sigma)!}
\right]^\half 
\left[ \frac{2 \lp+1}{4 \pi} \right]^\half 
\left[ \frac{2 \lp'+1}{4 \pi} \right]^\half \\
&~&
P^{\mu_p}_{\lp} (\cos \Theta) \,\,
P^{\mu_p}_{\lp'} (\cos \Theta) \,\,
\xi_{J,M,\eta,\Pi}(q;p,h) \, \,
\xi^{\dagger}_{J',M',\eta,\Pi'}(q;p',h) \,
\delta (\epsilon_p - \epsilon_h - \omega) \, . 
\label{wpm}
\end{eqnarray}
In the above expression we have defined
\begin{equation}
\xi_{J,M,\eta,\Pi}(q,\omega;p,h)=
\langle \Psi;JM;\Pi  (\lp \jp \emp \epsilon_p,\lh \jh \mh \epsilon_h) |\, 
J_{\eta,JM}(q,\omega) \, 
|\Psi;00;+1\rangle
\label{xi}
\end{equation}
and we made explicit the dependence on the associate Legendre
polynomials $P^\mu_l$.

With the above definitions
the cross section describing the nucleon emission induced by the
absorption of a photon can be written as \cite{gia85}
\begin{equation}
\frac {{\rm d} \sigma(\omega)}{{\rm d} \Omega} = 
\frac{2 \pi^2 e^2}{\omega} \,\,
\frac {|{\bf p}| M}{(2 \pi)^3} \,\, R_T(q=\omega,\omega) \, ,
\label{gpxsect}
\end{equation}
where $M$ is the emitted nucleon rest mass.

In the inclusive total photoabsorption cross section the emitted
nucleon is not detected. The cross section for this process is
obtained from the expression (\ref{gpxsect}) by summing all the
emission channels and integrating on all the possible directions of
the emitted nucleon \cite{def66}
\begin{eqnarray}
\sigma(\omega) &=& \sum_{\lp \jp \emp} \sum_{ \lh \jh \mh} \nonumber
\int {\rm d} \Omega \frac { {\rm d}\sigma(\omega)}{{\rm d} \Omega}\\
& =&
\frac{ 8 \pi^3 e^2} {\omega} \, \sum_{\eta,J,M,\Pi} \sum_{p,h} \,
|\xi_{J,M,\eta,\Pi}(q=\omega,\omega;p,h) |^2 \,
\delta (\epsilon_p - \epsilon_h - \omega) \, .
\end{eqnarray}
The last equation shows that, in the total photoabsorption
cross section, the nucleus is considered to make a transition from its
ground state to an excited state with good angular momentum and
parity. The interference between the excitation multipoles in 
Eq. (\ref{wpm}) vanishes because of the 
integration on $\theta_p$ and the orthogonality of the Legendre
polynomials. 

In Eq. (\ref{xi}) we have introduced the multipole 
operator $J_{\eta,JM}$, which comes from the corresponding 
current expansion and which is conveniently described
as sum of the electromagnetic operators 
$ T^{\rm E}_{JM}(q)$ and $
T^{\rm M}_{JM}(q)$ which generate, respectively, electric and magnetic
excitations \cite{ama93,ama94}. These operators are defined as
\begin{equation}
T^{\rm E}_{JM}(q) \, = \, \frac{1}{\omega} \,
\int {\rm d}^3r \, \left\{ \nabla \times 
\left[ j_J(q r) \, {\bf Y}^M_{JJ}(\hat{\br}) \right ] \right\} 
\cdot {\bf J} (\br) 
\label{tejm}
\end{equation}
and
\begin{equation}
T^{\rm M}_{JM}(q) \, = \,
\int {\rm d}^3r \, j_J(q r) \,{\bf Y}^M_{JJ}(\hat{\br}) \cdot 
{\bf J} (\br) \, ,
\label{tmjm} 
\end{equation}
where we have indicated with $j_J(q r)$ the spherical Bessel
function, with ${\bf Y}^M_{JJ}$ the vector spherical harmonics
\cite{edm57}, and with ${\bf J} (\br)$ the electromagnetic current
operator. We suppose that this operator is composed by the sum of
one- and two-body operators, these last ones produced by the exchange
of mesons.

The OB current is the sum of the convection
\begin{equation}
{\bf j}^c (\br) \,= \, \sum^A_{k=1} \, \frac{1}{i2M_k} \,
\frac{1+\tau_3(k)}{2} \,  \left[ \delta(\br -\br_k) \, \nabla_k \, 
+ \, \nabla_k \, \delta(\br -\br_k) \right] 
\label{conv}
\end{equation}
and magnetization
\begin{equation}
{\bf j}^m (\br) \,= \, \sum^A_{k=1} \, \frac{1}{M_k} \,
\left(\mu_P \frac{1+\tau_3(k)}{2} + \mu_N \frac {1 - \tau_3(k)}{2} \right) 
\, \nabla_k \times \delta(\br -\br_k) \, \bsigma(k) 
\label{mag}
\end{equation}
currents. In the above equations $M_k$ indicates the mass of the $k$-th
nucleon, $\mu_P$ and $\mu_N$ the anomalous magnetic moments of the
proton and neutron respectively, $\bsigma(k)$  the Pauli matrices 
and $\tau_3(k)=1$ if the $k$-th nucleon is a proton and  $\tau_3(k)=-1$
in case it is a neutron.

The above expressions refer to point-like nucleons. In
actual calculations we folded them with the electric and magnetic nucleon
form factors, $G_{\rm E,M}^{\rm P,N}(q,\omega)$. We used the parameterization 
of Ref. \cite{hoe76}.

To give the expressions of the MEC it is convenient to define the
function $h(\br)$ as the Fourier transform of the dynamical pion
propagator \cite{ama94}
\begin{equation}
\label{hache}
h(\br-\br_l)=\int \frac{{\rm d}^3k}{(2\pi)^3}\,
\frac{F_{\pi\rm N}(k,\varepsilon)\, 
\e{i\bkey \cdot (\br-\br_l)}} 
{k^2+m_{\pi}^2-\varepsilon^2} \, ,
\end{equation}
where we have indicated with $F_{\pi\rm N}$ the pion--nucleon form
factor and with $\varepsilon=(\Delta E)_l$ the energy of the exchanged
pion, of mass $m_{\pi}$, obtained as the difference between the
energies of the final and initial single particle states of the $l$-th
nucleon.

%%%%%%%%%%%%%%%%%%%%%%%%%%%%%%%%%%%%%%%%%%%%%%%%%%%%%%%%%%%%
\begin{figure}
\vspace*{-3cm}
\hspace*{.7cm}
\includegraphics[bb=50 50 500 700,angle=0,scale=0.8]
       {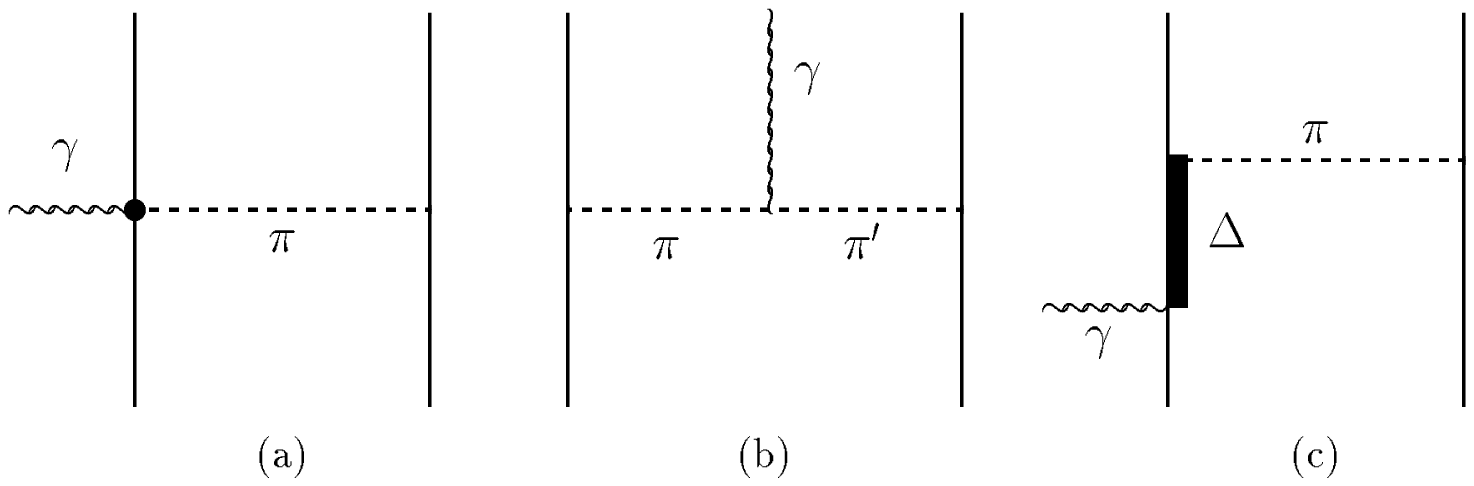}
\vspace*{-11.cm}
\caption{\small Feynman diagrams of the three MEC operators we
  consider: (a) seagull, (b) pionic and (c) $\Delta$ isobar.
}
\label{fig:fey}
\end{figure}
%%%%%%%%%%%%%%%%%%%%%%%%%%%%%%%%%%%%%%%%%%%%%%%%%%%%%%%%%%%%

In our work we consider the three types of MEC represented by  the
Feynman diagrams of Fig. \ref{fig:fey}. They are the seagull (a), the
pionic (b) and the virtual $\Delta$ excitation (c) terms. The
expressions of the seagull and pionic currents are,
respectively,
\begin{equation} 
\label{seagull}
{\bf j}^{\rm S} ({\bf q},\omega) =
4\pi\frac{f_{\pi}^2}{m_{\pi}^2} \, F_{\rm S}(q,\omega)
\sum_{k,l=1 \atop k\ne l}^A \,
[\btau (k) \times \btau (l)]_{_3} \, 
\e{i\bqu \cdot \br_{k}}
 \bsigma (k)  \, 
 \bsigma (l)  \cdot\nabla_k h({\bf r}_k-{\bf r}_l)
\, ,
\end{equation}
and
\begin{eqnarray}
\nonumber
{\bf j}^{\pi}({\bf q},\omega) & =& - 4\pi\frac{f_{\pi}^2}{m_{\pi}^2} \,
F_{\pi}(q,\omega) \,
\sum_{k,l=1 \atop k\ne l}^A \,
[\btau (k) \times \btau (l)]_3 \,\,
 \int {\rm d}^3r\,\e{i\bqu\cdot\br}\, 
\\ \label{pionic} & &
 \bsigma (k)  \cdot \nabla h({\bf r}-{\bf r}_k)\, 
 \nabla \, \left[
 \bsigma (l) \cdot \nabla h({\bf r-r}_l)
\right] \, .
\end{eqnarray}

The situation for the $\Delta$--isobar current is not so well defined
because formulations based upon static quark models \cite{che71,hoc73} or
chiral lagrangian \cite{pec69,ord81} give different expressions. We have
adopted the point of view of the first group of authors and use the
expression
\begin{eqnarray} 
\nonumber
{\bf j}^{\Delta}({\bf q},\omega) &= & -i C_{\Delta} F_{\Delta}(q,\omega) 
\sum_{k,l=1 \atop k\ne l}^A \,
\e{i\bqu\cdot\br_k}\bqu 
\\ & & \nonumber
\displaystyle \left\{ 
[ \btau (k) \times \btau (l)]_3 \,
 \bsigma (k) \times \nabla_k \, 
 \bsigma (l) \cdot \nabla_k \, h({\bf r}_k-{\bf r}_l) 
 \, - \, 4 \tau^3 (l) \, \nabla_k \, 
 \bsigma (l) \cdot \nabla_k \, h({\bf r}_k-{\bf r}_l)
\right\} \, .
\label{isobar}
\end{eqnarray}
In the above equations 
$f_{\pi}=0.079$ is the effective pion-nucleon coupling constant and
\begin{equation}
\label{cdelta}
C_{\Delta} = 4\pi\frac{f_{\pi}^2}{m_{\pi}^2}
          \,   \frac{4}{25M(M_{\Delta}-M)} \, ,
\end{equation}
with $M_{\Delta}$ the $\Delta$ mass. We have indicated with the
symbols $F_\pi$, $F_\Delta$ the electromagnetic form factors of the
pion and of the $\Delta$ and with $F_S$ the electric isovector
nucleon form factor. To be consistent with the one-body currents we
used the following expressions for $F_S$ and $F_\pi$
\begin{eqnarray}
F_{\rm S}(q,\omega) & = & 
G_{\rm E}^{\rm P}(q,\omega) - G_{\rm E}^{\rm N}(q,\omega) ,\\
F_{\pi}(q,\omega)   & = & F_{\pi\gamma}(q,\omega) = 
\frac{1}{1+(q^2-\omega^2)/m_{\rho}^2}
\, ,
\end{eqnarray}
where $m_{\rho}$ is the mass of the $\rho$-meson. 

The situation is more complicated for the $\Delta$ current since the
electromagnetic form factor $F_\Delta$, and the constant $C_{\Delta}$
are model dependent. The major uncertainty is related to $C_{\Delta}$,
but a discussion of this problem is beyond the aim of the present
work. The expression (\ref{cdelta}) of $C_{\Delta}$ we have chosen is
widely used in the literature \cite{hoc73,ris84}. For the form factor
$F_{\Delta}$, following the static quark model, we have used
\begin{equation}
F_{\Delta}(q,\omega) \, =  \, 
2G_{\rm M}^V (q,\omega) \, = \,
G_{\rm M}^{\rm P}(q,\omega) - G_{\rm M}^{\rm N}(q,\omega) \, .
\end{equation}
Finally, we would like to comment on the pion-nucleon form factor
\begin{equation}
F_{\pi\rm N}(k,\varepsilon) =
\frac{\Lambda^2-m_{\pi}^2}{\Lambda^2+k^2-\varepsilon^2}
\end{equation}
included in the expression (\ref{hache}) of the pion propagator. We
have verified \cite{ama93} that, in the quasi-elastic peak region, for
the values of $\Lambda$ commonly accepted ($\sim 1$~GeV), the results
are very close to those obtained considering simply $F_{\pi\rm N} =1$,
which is the value we have adopted.

\section{THE NUCLEAR MODEL}
\label{nucmod}
In the previous section we have described 
the reaction mechanism but
we did not specify the structure of the
nuclear ground excited states.
These are the two inputs required to calculate the $\xi$
functions in Eq. (\ref{xi}). 

Because of energy conservation, the knowledge of the energy, and of
the momentum, of the emitted nucleon identifies the quantum numbers of
the hole state of the residual nucleus.  The state $|\Psi; J M; \Pi;
(\lp \jp \emp \epsilon_p,\lh \jh \mh \epsilon_h) \rangle$ is
asymptotically characterized by the quantum numbers of a 
1p-1h excitation we label generically with $p$ and $h$.

In our model the nuclear states are described as
\begin{eqnarray}
\label{jasti}
|\Psi; 0 0; +1\rangle &=&
\frac{ F \,\, |\Phi;0 0; +1\rangle } 
{\langle \Phi;0 0; +1|\,\, F^\dagger F \,\, |\Phi;0 0; +1\rangle ^\half}
\,\,, \\
\label{jastf}
|\Psi; J M; \Pi;p,h\rangle &=& 
\frac {F \,\, |\Phi;J M; \Pi;p,h\rangle}
{\langle \Phi;J M; \Pi;p,h| \,\, F^\dagger F \,\, |\Phi;J M;
  \Pi;p,h\rangle^\half }
 \, .
\end{eqnarray}
We have indicated with $|\Phi;0 0; +1\rangle$ the Slater determinant
describing the mean field wave function in a pure 
IPM. This means that, given a basis of single particle wave
functions, all the states below the Fermi surface are fully occupied
and those above are all completely empty.  The state $|\Phi;J M; \Pi;p
h\rangle$ indicates a Slater determinant where the hole function $h$
has been substituted with the continuum particle function $p$.  With
respect to the IPM, the novelty of our ansatz is the presence of the
correlation function $F$.  This function has, in principle, a very
complicated operatorial dependence, analogous to that of the
hamiltonian \cite{fab00,fab98}. In this work we restrict our
calculations to the case of purely scalar (Jastrow) correlations,
therefore we immediately simplify the expressions formulating them
only in terms of this type of correlations. The adopted ansatz about 
the correlation is \cite{jas55}
\beq
F(1,2,...A)=\prod^A_{i<j} f(r_{ij}) \, ,
\label{corf}
\eeq
where $r_{ij}=|{\bf r}_i-{\bf r}_j|$ is the distance between the
the particles $i$ and $j$.

In order to make use of well-established cluster expansion techniques
\cite{fan87} we rewrite the transition matrix
$\xi_{J,M,\eta,\Pi}(q,\omega=\epsilon_p - \epsilon_h;p,h)$ 
in Eq. (\ref{xi}) as
\begin{eqnarray}
\nonumber
\xi_{J,M,\eta,\Pi}(q,\epsilon_p - \epsilon_h;p,h)  &=& 
\frac {\langle \Psi;  J M; \Pi;p,h|  J_{\eta,JM}(q) 
                    |\Psi; 0 0; +1 \rangle} 
      {\langle \Psi; 0 0; +1 |\Psi; 0 0; +1 \rangle} \\
  &~&    \left[ \frac { \langle \Psi;  0 0; +1 |\Psi; 0 0; +1 \rangle}
               { \langle \Psi;  J M; \Pi;p,h|\Psi; J M; \Pi;p,h\rangle}
      \right]^\half \, .
\label{xi2}
\end{eqnarray}
The two factors in Eq. (\ref{xi2}) are separately evaluated by
expanding both numerator and denominator in powers of the short-range
correlation function. The presence of the denominators is used to 
eliminate the unlinked diagrams \cite{fan87}.

The matrix element to be calculated in Eq. (\ref{xi2}) is
\begin{eqnarray}
\nonumber
&~&\hspace*{-3cm} \langle \Psi;J M; \Pi;p,h| J_{\eta,JM}(q) 
|\Psi; 0 0; +1 \rangle \\
\nonumber
&=&
\langle\Phi;J M; \Pi;p,h| F^\dagger 
 J_{\eta,JM}(q) F |\Phi; 0 0; +1 \rangle_L \\
\nonumber
&=&  
\langle\Phi;J M; \Pi;p,h| 
       J_{\eta,JM}(q) \prod^A_{i<j}f^2(r_{ij}) |\Phi; 0 0; +1 \rangle_L \\
&=& 
 \langle\Phi;J M; \Pi;p,h| 
 J_{\eta,JM}(q) \prod^A_{i<j}(1+h_{ij})|\Phi; 0 0; +1 \rangle_L \, ,
\end{eqnarray}
where we have used the function $h_{ij}=f^2(r_{ij})-1$ and
the subindex $L$ indicates that only the linked diagrams are
considered. 

Up to this moment the treatment of the transition matrix element is
the same as that adopted in nuclear matter \cite{fan87,fab89}. We
insert at this point a new approximation consisting in retaining only
those terms where the $h_{ij}$ function appears only once
\begin{eqnarray}
\nonumber
\xi_{J,M,\eta,\Pi}(q,\epsilon_p - \epsilon_h;p,h) & \longrightarrow &
\xi^1_{J,M,\eta,\Pi}(q,\epsilon_p - \epsilon_h;p,h) \\
&\,=\,& \langle \Phi;J M; \Pi;p,h | \,J_{\eta,JM}(q)\, 
\sum_{i<j}\, (1+h_{ij}) \,|\Phi; 0 0; +1  \rangle_L \, .
\label{ximodel}
\end{eqnarray}
This result has been obtained using a procedure analogous to that
adopted in Ref. \cite{co95} to evaluate the density
distribution, therefore the truncation of the expansion is done only
after the elimination of the unlinked diagrams.

By identifying with $1$ the coordinate of the nucleon struck by the
photon, we can express the $\xi^1$ in Eq. (\ref{ximodel}) as
\begin{eqnarray}
\nonumber
\xi^1_{J,M,\eta,\Pi}(q,\epsilon_p - \epsilon_h;p,h) \, &=&\, 
\langle \Phi;J M; \Pi;p,h | \, J_{\eta}(q)\,|\Phi; 0 0; +1  \rangle \\
\nonumber
\,&~& + \,  \langle \Phi;J M; \Pi;p,h |
        \, J_{\eta}(q) \, \sum^{A}_{j>1}\, h_{1j}\, 
          |\Phi; 0 0; +1  \rangle_L     \\
\, &~& + \,  \langle \Phi;J M; \Pi;p,h |
\, J_{\eta}(q) \, \sum^{A}_{1< i <j } \, h_{ij} \, 
 |\Phi; 0 0; +1  \rangle_L \, .
\label{xi1p1h}
\end{eqnarray}
%

%%%%%%%%%%%%%%%%%%%%%%%%%%%%%%%%%%%%%%%%%%%%%%%%%%%%%%%%%%%%
\begin{figure}
\hspace*{1cm}
\includegraphics[bb=50 50 500 700,angle=0,scale=0.7]
       {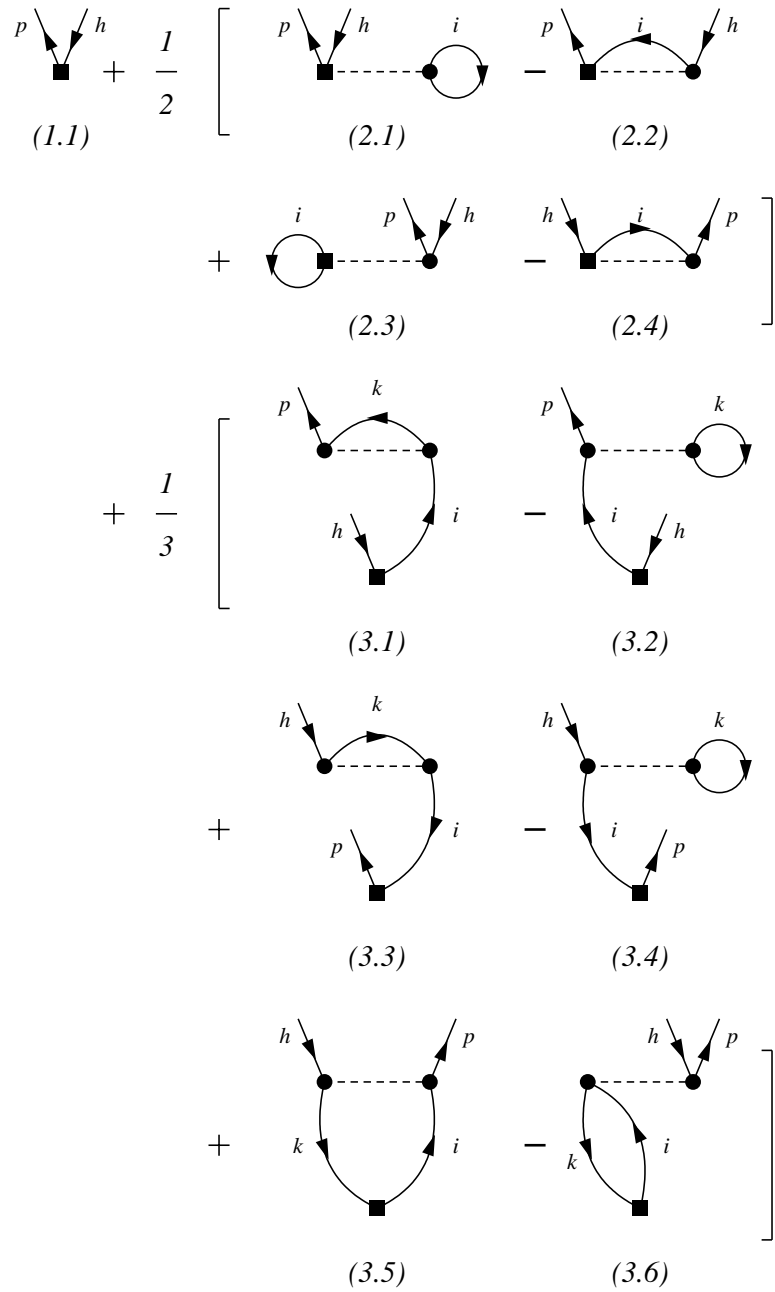}
\caption{\small Mayer-like diagrams describing the SRC
             considered in our calculations. 
}
\label{fig:mey}
\end{figure}
%%%%%%%%%%%%%%%%%%%%%%%%%%%%%%%%%%%%%%%%%%%%%%%%%%%%%%%%%%%%

The various terms generated by the above expression are shown as
Mayer-like diagrams 
\cite{may40} in Fig. \ref{fig:mey}.  The first term in Eq.
(\ref{xi1p1h}) is the IPM contribution and it corresponds to the
one-point diagram (1.1) of Fig. \ref{fig:mey}. The other two terms in
Eq. (\ref{xi1p1h}) are represented by the two- and three-point
diagrams.  The presence of these last diagrams is necessary to have
the correct normalization of the many-body wave function, as discussed
in \cite{co01}.

What should be specified now is the set of single particle
wave functions and the correlation functions. We took them from the
FHNC calculations of Refs. \cite{ari96,fab00} where these
ingredients of the calculation have been fixed by minimizing the
ground state expectation value of the hamiltonian. In
Ref. \cite{ari96} the set of 
single particle wave functions was taken
from the literature \cite{co85,co87} and the minimization was done
only by changing the correlation. This procedure was justified because
the nucleon--nucleon interaction was semi-realistic. 
In the calculations of the present work we have used the correlations
obtained with the S3 interaction of Afnan and Tang \cite{afn68} by means
of the Euler and gaussian procedures (see Ref. \cite{ari96}).
For sake of comparison in our calculations we have also used the
scalar part of the state dependent correlation function fixed in
Ref. \cite{fab00}.

The calculations of the transition matrix elements are carried out by
performing a multipole expansion of the correlation function $h_{ij}$
\cite{co95,co01,mok01}. The single particle wave functions are
described in a spherical basis, and we used the traditional angular
momentum coupling techniques to evaluate the matrix elements.  The
explicit expressions of these matrix elements for the magnetization
current are given in Ref. \cite{co01}. Those of the convection
current are given in Ref. \cite{mok01a}.

The MEC have have been calculated only in the IPM, i. e. neglecting
the last two terms in Eq. (\ref{xi1p1h}). Their expressions are given
in Refs. \cite{ama93,ama94,ama93a}. As already stated in the introduction,
in our computational scheme SRC and MEC are related only by the
interference terms of the transition matrix elements.

\section{SPECIFIC APPLICATIONS}
\label{applications}

The nuclear model above presented has been applied to describe nucleon
emission from the $^{16}$O nucleus. The motivations of this
choice have already been outlined in the introduction. It is a
spherical doubly-closed shell nucleus and the implied symmetries
simplify the treatment of shell and finite size effects. 
For these reasons the
ground state of this nucleus has been the subject of investigation of
microscopic theories \cite{pie92}-\cite{fab00}. Also our calculations
relative to inclusive \cite{co01} and exclusive \cite{mok01} electron
scattering experiments have been done for $^{16}$O. By studying this
nucleus we can therefore make a comparison with our previous results
and also test the consistency of the description of the various
experiments.  Obviously the formalism could have been applied to
study any other doubly closed shell nucleus.

\subsection{Testing the direct knock-out model}
\label{rpa}

Our model has been developed with the aim of providing a good
description of the SRC correlations. Since the presence of collective
phenomena has not been considered, the range of applicability of our
model is limited to those kinematic situations where the effects of
these phenomena can be neglected.

In our previous works the model has been applied in the quasi-elastic
region \cite{co98,ama98,co01,mok01}. It has been verified in Ref.
\cite{co88} that in this region collective excitations described in
terms of continuum RPA are negligible and their effects become smaller
with increasing momentum transfer. The modification of the IPM
responses is produced by the Final State Interactions (FSI) between
the emitted nucleon and the rest nucleus.  Second RPA calculations
\cite{dro87} showed that, in the quasi-elastic region, the FSI effects
can be accounted for by using an optical potential.  The mean-field
model implemented with optical potentials taken from nucleon-nucleus
elastic scattering data is the approach proposed by the Pavia group
\cite{bof96,bof93} and successfully used to analyze the nucleon
emission data.

In Ref. \cite{mok00} our model was applied to study low-lying states
excitations.  Also in this case we excluded collective excitations and
we studied data at large values of the momentum transfer, above 300
MeV/c.

Clearly, the electron scattering kinematics ensuring the dominance of
the single particle excitations are not achievable with photons.
Furthermore, in the literature it is well known that continuum RPA
calculations successfully describe photon reactions even at energies
above the giant resonance region \cite{gia85}.  The photon excitation
of doubly closed shell nuclei has been widely studied with
self-consistent continuum RPA with Skyrme interactions, first by the
Bologna \cite{cav82}-\cite{sar93} and later by the Gent group
\cite{ryc87}-\cite{ryc88}. For this reason we felt necessary to study
the importance of collective excitation modes in photo-reactions.

Our study of the effects of collective excitations 
has been done by using the Fourier-Bessel Continuum
RPA approach of Ref. \cite{deh82} used in the past to investigate
coincidence electron scattering processes in the giant resonance
region \cite{co85,co87} and inclusive quasi-elastic responses
\cite{co88}.  This RPA approach is based upon the Landau-Migdal theory
of finite Fermi systems.  The ground state is
described by a phenomenological mean field potential and the residual
effective interaction is fixed to reproduce some general properties of
the system.

In our case the single particle basis has been generated by a mean
field potential of Woods-Saxon type. In our RPA calculations both
bound and continuum single particle wave functions have been produced
by the same real Woods-Saxon potential. As already discussed in the
previous section, the parameters of the potential are those used in
Ref. \cite{ari96} for the FHNC calculation. 

The residual interaction has the form
\begin{eqnarray}
\nonumber
V(1,2) &=& C_0 \left[
     f(r_{12}) + f'(r_{12}) \bsigma(1) \cdot \bsigma(2) 
      \right. \\ 
     &+& g(r_{12}) \btau(1) \cdot \btau(2) 
     + g'(r_{12})  \bsigma(1) \cdot \bsigma(2) \btau(1) \cdot \btau(2) 
     \left. \right] \, ,
\label{interaction}
\end{eqnarray}
where we have indicated with $r_{12}$ the distance between the two
interacting nucleons.  In the above equation we used the traditional
nomenclature to identify the various terms of the force. One should
not confuse the  $f(r_{12})$  in Eq. (\ref{interaction}) with the
correlation function in Eq. (\ref{corf}).
We made calculations with two types of
effective forces: a zero-range Landau-Migdal interaction and the
finite range polarization potential of Ref. \cite{pin88}.
For the Landau-Migdal interaction the functions $f$, $f'$, $g$, 
and $g'$
are constants. The values of these parameters have been chosen as in
Ref. \cite{co85} where it has been shown that they provide a
reasonable description of the giant resonances and of the low-lying
excited states.  With this set of values we obtained the energy of the
collective low-lying 3$^-$ state at 6.40 MeV against the experimental
value of 6.15 MeV.

One of the interesting features of the finite range polarization
potential is the fact that it is built to take into account
effectively the exchange terms.  This fact is very important in our case,
since our Fourier-Bessel RPA calculations do not consider these terms.
We straightforwardly used the interaction as given in Ref. \cite{pin88}
and we obtained for the energy of the low-lying 3$^-$ state the value
of 6.65 MeV.

\begin{figure}
%\vspace*{-4.cm}
\includegraphics[bb=50 50 500 700,angle=90,scale=0.8]
       {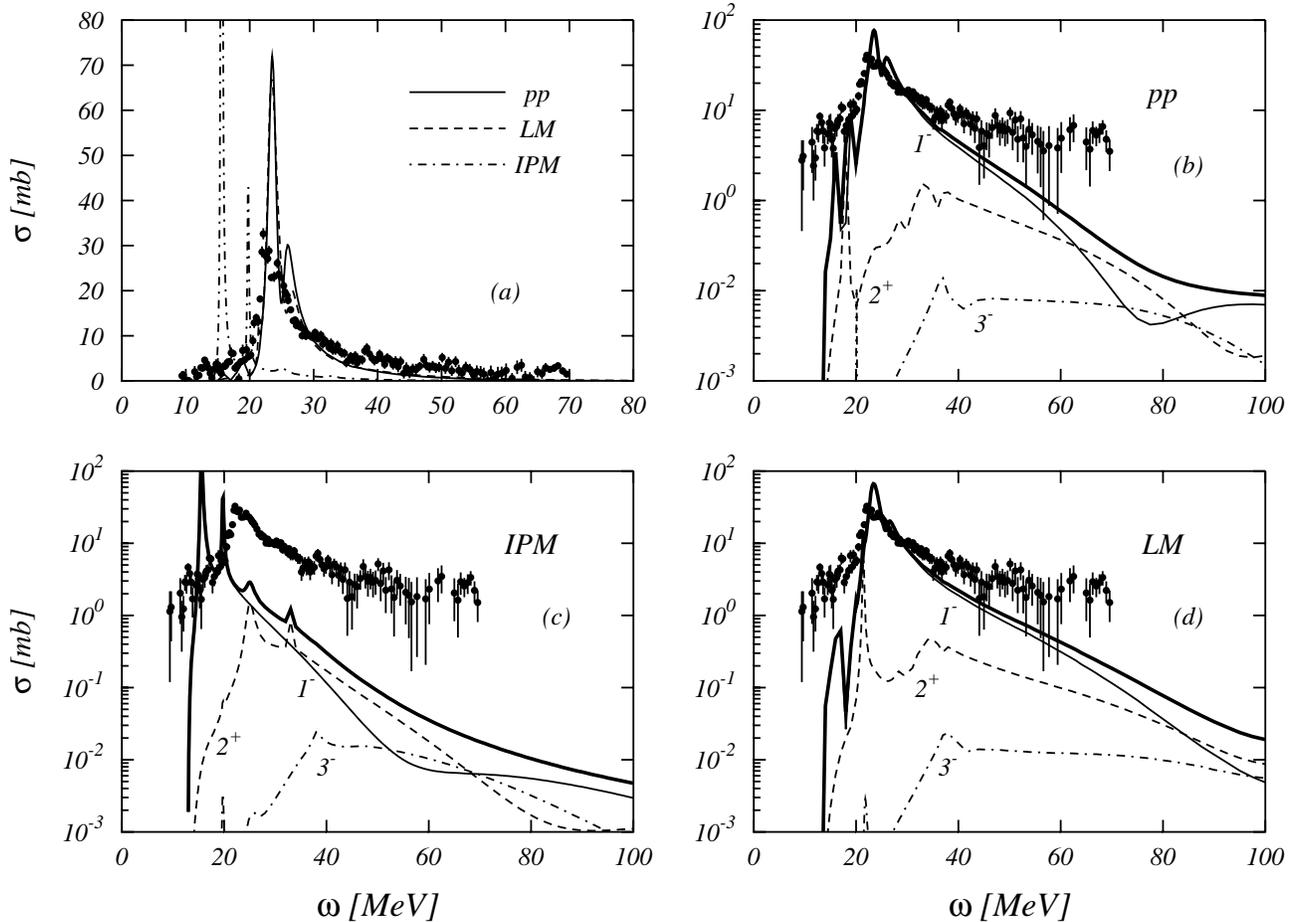}
\vspace*{-1.5cm}
\caption{\small Total photoabsorption cross sections calculated in the
  IPM and with the continuum
  RPA by using the polarization
  potential (pp) and the Landau-Migdal (LM) residual interactions. In 
  panels $(b)$, $(c)$ and $(d)$ the contribution of the main electric
  multipoles is shown and the full thick line represents the total
  cross section. The data are from Ref. \protect\cite{ahr75}.
}
\label{fig:phtot}
\end{figure}

All the calculations we shall discuss in this subsection have been
done by using OB currents only. In Fig. \ref{fig:phtot} we compare our
results, obtained by considering all the excitation multipoles up to
$J=3$, with the total photoabsorption data of Ref. \cite{ahr75}.
The panel $(a)$ of the figure shows that
while the RPA calculations are able to reproduce the
centroid energy of the giant resonance, the IPM cannot do it. However,
even in RPA results, the height of the peak is overestimated and the
width underestimated.  
These facts are rather well known in the literature
\cite{spe91}.
The deficiencies of the RPA have been widely
investigated, and it seems that 
they can be cured by considering excitations beyond
1p-1h, like in the Second RPA \cite{dro90} or in the phonon coupling
scheme \cite{ber83,kam96}.

%%%%%%%%%%%%%%%%%%%%%%%%%%%%%%%%%%%%%%%%%%%%%%%%%%%%%%%%%%%%%%%%%%%%%%
%%%%%%%%%%%%%%%%%%%%%%%%%%%%%%%%%%%%%%%%%%%%%%%%%%%%%%%%%%%%%%%%%

Since we are interested in the region beyond the giant resonance, we
emphasize the comparison between theory and experiment in the panels
$(b)$, $(c)$ and $(d)$ of Fig. \ref{fig:phtot} by using the logarithmic
scale. In these three panels we show the contribution to the total
cross sections of each electric multipole. The contribution of the
magnetic multipoles is too small to appear in the figure.  The
three calculations show that the 1$^-$ excitation is the dominant one
in the giant resonance region, but with increasing energy the
contribution of the 2$^+$, and that of the 3$^-$, becomes important.
The relevance of the high multipolarity is related to the residual
interaction. In the IPM the 2$^+$ cannot be neglected already at 40
MeV, while the dominance of the 1$^-$ is extended at larger energies
in the RPA calculations: 60 MeV for the polarization potential 
and 80 MeV for the LM interaction.  
This result is in agreement with the findings of Ref.
\cite{ryc88}.  The IPM results are, in the region above 40 MeV, one
order of magnitude smaller than those obtained with the RPA.  One
should remark, however, that even the RPA calculations are below the
data in the region of interest.

%%%%%%%%%%%%%%%%%%%%%%%%%%%%%%%%%%%%%%%%%%%%%%%%%%%%%%%%%%%%%%%%%%%%%%%%
\begin{figure}
\hspace*{.5cm}
\includegraphics[bb=50 50 500 680,angle=90,scale=0.8]
               {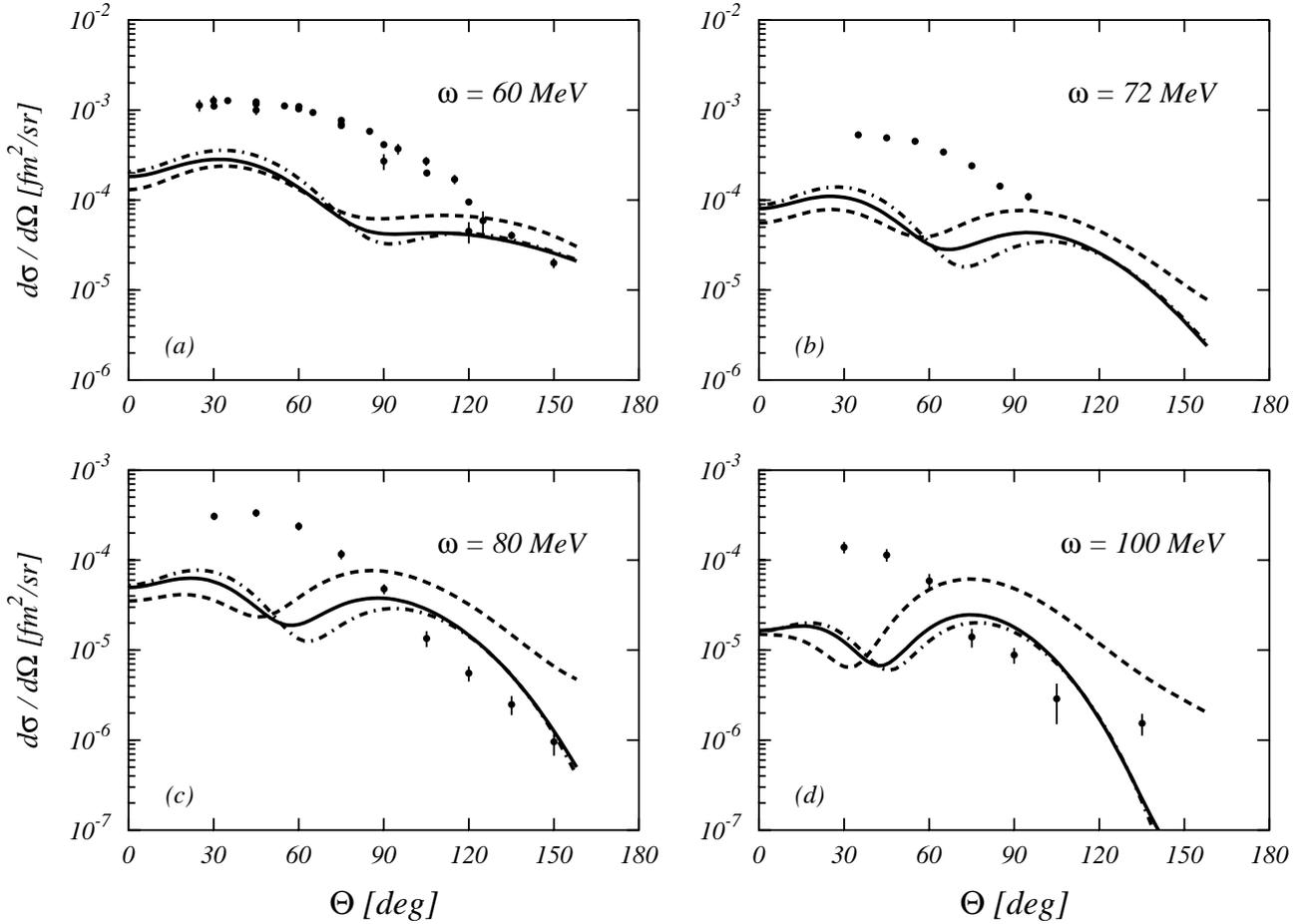}
\vspace*{-1.5cm}
\caption{\small  Angular distributions of the
  $^{16}$O($\gamma$,p$_0$)$^{15}$N reaction 
  for various energies of the photon. 
  The full lines show the RPA results obtained with the polarization
  potential, the dashed lines the RPA results obtained with Landau-Migdal
  interaction and the dashed-dotted line the IPM results. The data
  have been taken from Ref. \protect\cite{fin77} and also from
  \protect\cite{des93,mil95} in the panel $(a)$.
  }
\label{fig:gprpa}
\end{figure}
%%%%%%%%%%%%%%%%%%%%%%%%%%%%%%%%%%%%%%%%%%%%%%%%%%%%%%%%%%%%%%%%%%

\begin{figure}
\vspace*{-3.5cm}
\includegraphics[bb=50 50 500 700,angle=90,scale=0.8]
               {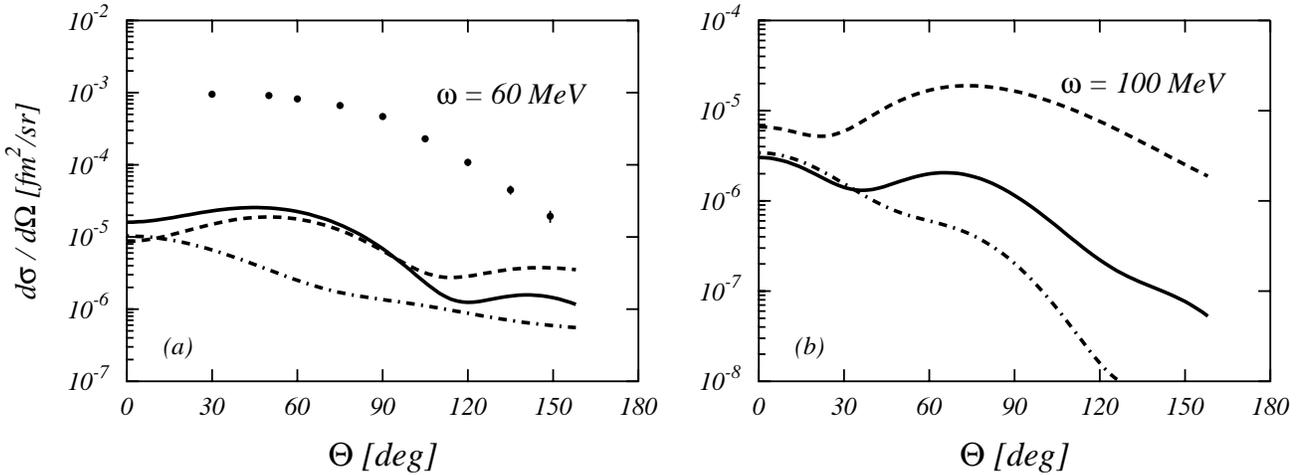}
\vspace{-3.0cm}
\vspace*{-1.5cm}
\caption{\small Angular distribution of the 
  $^{16}$O($\gamma$,n$_0$)$^{15}$O reaction for two photon energies.
  The meaning of the lines is the same as in Fig. \protect\ref{fig:gprpa}.
  The data are from Ref. \protect\cite{gor82}.
  }
\label{fig:gnrpa}
\end{figure}

If the total photoabsorption cross section cannot be reproduced we
have little hope to reproduce the exclusive cross section. Indeed the
observation of Figs. \ref{fig:gprpa} and \ref{fig:gnrpa}
confirms this expectation. In these
figures we compare the $^{16}$O($\gamma$,p$_0$)$^{15}$N and
$^{16}$O($\gamma$,n$_0$)$^{15}$O data of Refs.
\cite{fin77}-\cite{mil95} and \cite{gor82} with our RPA and IPM
calculations. All the curves underestimate the data, as we expected.

%%%%%%%%%%%%%%%%%%%%%%%%%%%%%%%%%%%%%%%%%%%%%%%%%%%%%%%%%%%%%%%%%%
%%%%%%%%%%%%%%%%%%%%%%%%%%%%%%%%%%%%%%%%%%%%%%%%%%%%%%%%%%%%%%%%%%

In Fig. \ref{fig:gprpa} the
IPM and the RPA results obtained with the polarization potential are
rather similar, while those obtained with the Landau-Migdal
interaction differ more. This fact is consistent with the study done
in the electron scattering case \cite{co88,bub91}. 
The finite range interactions become weaker with increasing momentum,
while contact interactions are constant in momentum space. For this
reason the zero-range
interaction, tuned to reproduce low-energy nuclear properties,
overestimates the role of the RPA above the giant resonance region.
The momentum behavior of 
finite-range interactions is more realistic
than that of
the zero-range forces, therefore in the estimate of the collective 
effects we refer to the results obtained with the polarization
potential. These results indicate that
the RPA effects become smaller with the increase of the
photon energy. For proton emission at 100 MeV these effects are really
negligible, and the cross section is essentially that of the IPM.

The situation for the neutron emission is rather different, as it is
shown in Fig. \ref{fig:gnrpa}. Here IPM and RPA results are very
different. The photon couples to a neutron only through its magnetic
moment and, only for the 1$^-$ excitation to its effective charge
\cite{bla52}. In sect. \ref{mec} we shall show that, at the photon
point, the magnetization current is much smaller than the convection
one. For this reason the dominant mechanism ruling the neutron
emission is not the direct knock-out, but a re-scattering process
where the proton struck by the photon interacts with the neutron which
is emitted. Continuum RPA calculations partially take into account
these re-scattering phenomena, for this reason RPA neutron emission
cross sections are much larger than the IPM cross sections. This
increase is however not sufficient to obtain a reasonable description
of the 60 MeV data of panel $(a)$.

The fact that experimentally the neutron emission cross section is of
the same order of magnitude of the proton emission cross section is an
open 
problem deserving further investigation, but it is out of the scope of
the present paper. Our RPA calculations were aimed to test the
applicability of the direct knock-out model. 
From the results we have presented 
we may conclude that this
model can be used to describe the proton photo-emission for energies
above the giant resonance region. Henceforth we shall restrict our
investigation to this case.

Even though we restrict ourselves to
the case of proton emission, where IPM and RPA results are
very similar, the comparison with the experimental data shown in
Fig. \ref{fig:gprpa} indicates that some important physics effects are
still missing.
We have
already mentioned the relevance of the FSI in the description of the
quasi-elastic electron scattering data. In the calculation of the 
inclusive responses
\cite{co01} we have treated the FSI using the folding model developed
in Ref. \cite{co88}. In treating (e,e'p) reactions we chose the
approach of the Pavia group \cite{bof96,bof93} which takes into account
the FSI
by describing the emitted nucleon as moving in an optical potential.
We have adopted the same strategy also in the present work.  In this
type of calculations the single particle basis is not any more
orthonormal since particle and hole wavefunctions are described by two
different mean fields. This problem has been investigated in Refs.
\cite{bof82,bof84} where it has been found that the effects of the 
non orthogonality of the basis are not important under in the 
kinematic conditions of interest.

%%%%%%%%%%%%%%%%%%%%%%%%%%%%%%%%%%%%%%%%%%%%%%%%%%%%%%%%%%%%%%%%%%%%%
\begin{figure}
\vspace*{-3cm}
\includegraphics[bb=50 50 500 700,angle=90,scale=0.8] 
{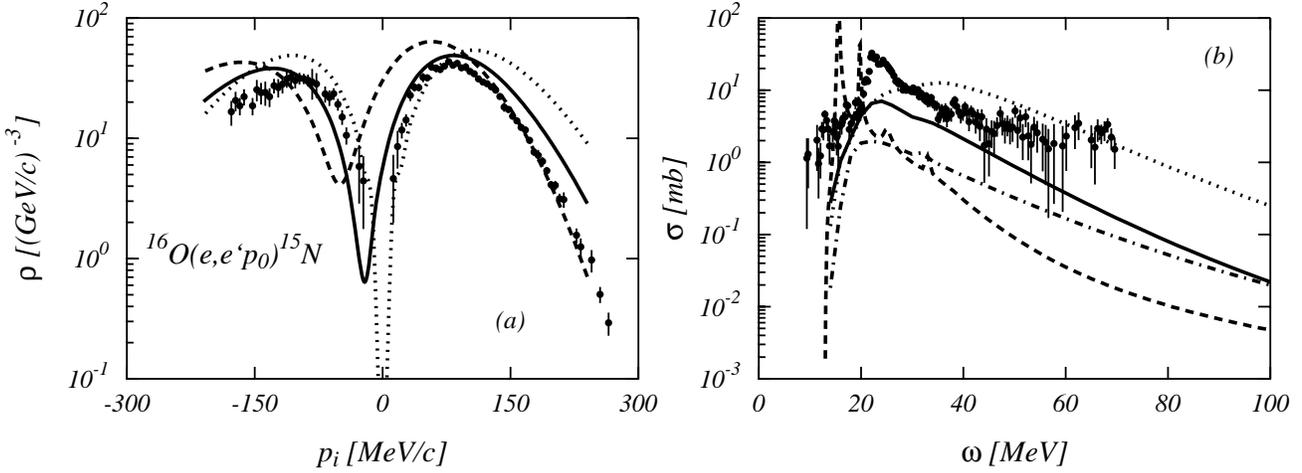}
\vspace {-4.3cm}
\caption{\small Panel $(a)$: reduced $^{16}$O(e,e'p$_0$)$^{15}$N 
  cross section compared with the data of Ref. \protect\cite{leu94}.
  The dashed line has been obtained with the real mean-field
  potential. The full line corresponds to the calculation performed 
  by using the optical potential
  of Ref. \protect\cite{sch82} for the particle states. The dotted
  line show the PWIA results.  Panel $(b)$:
  total photoabsorption cross section 
  (data of Ref. \protect\cite{ahr75}). The
  full, dashed and dotted lines have same meaning as those of panel
  $(a)$. The dashed dotted line has been calculated considering 
  the energy dependence of the optical potential. }
\label{fig:opt}
\end{figure}
%%%%%%%%%%%%%%%%%%%%%%%%%%%%%%%%%%%%%%%%%%%%%%%%%%%%%%%%%%%%%%%%%%%%

In our calculations we used the optical potential of Schwandt 
{\it et al.}
\cite{sch82} which was adopted in the analysis of NIKHEF 
$^{16}$O(e,e'p)$^{15}$N 
data \cite{leu94}.  In the panel $(a)$ of Fig.
\ref{fig:opt} we show the reduced cross sections of the above reaction
calculated with the real mean field potential (dashed line) and with
the optical potential (full line). The improvement in the description
of the data is evident, even though the theory is still above the
experimental points.  The dotted line show the result obtained by
considering the wave function of the emitted particle to be a plane
wave. This approximation is usually called Plane Wave Impulse
Approximation (PWIA) \cite{bof96}.

In the panel $(b)$ of the same figure we show the total photoabsorption
cross section. The full, dashed and dotted lines have been obtained
with the same inputs used for the analogous curves of panel $(a)$.  The
results of this figure indicate that, above the giant resonance
region, the cross section strongly depends on the mean--field
potential describing the emitted particle. The cross section becomes
smaller when the depth of the real part of this potential increases.
The lowest cross section (dashed line) is obtained by using the purely
real Woods-Saxon potential of Ref. \cite{ari96}
which is the deepest one among those we
have used. As expected, the dotted line showing the PWIA results is
above all, experimental data included.  In between there are the
results obtained with the Schwandt 
{\it et al.} potential. Specifically, the
full line has been obtained with the parameters fixed to evaluate the
(e,e'p) cross section of the panel $(a)$.  In this case the real part of
the potential is shallower than that of the potential of
Ref. \cite{ari96}.
The dashed-dotted line was calculated changing the parameters of the
optical potential at each excitation energy, strictly following the
parameterization  given in Ref. \cite{sch82}.
The real part of the potential becomes shallower with increasing 
energy. This effect is present in the phenomenological optical
potentials fixed to fit elastic nucleon-nucleus scattering data, like
the potential we are using, and also in the optical potentials
evaluated in microscopic many-body theories \cite{jeu76,fan83}.

%%%%%%%%%%%%%%%%%%%%%%%%%%%%%%%%%%%%%%%%%%%%%%%%%%%%%%%%%%%%%%
\begin{figure}
\vspace*{-5.cm}
\hspace*{.5cm}
\includegraphics[bb=50 50 600 680,angle=90,scale=0.8] 
{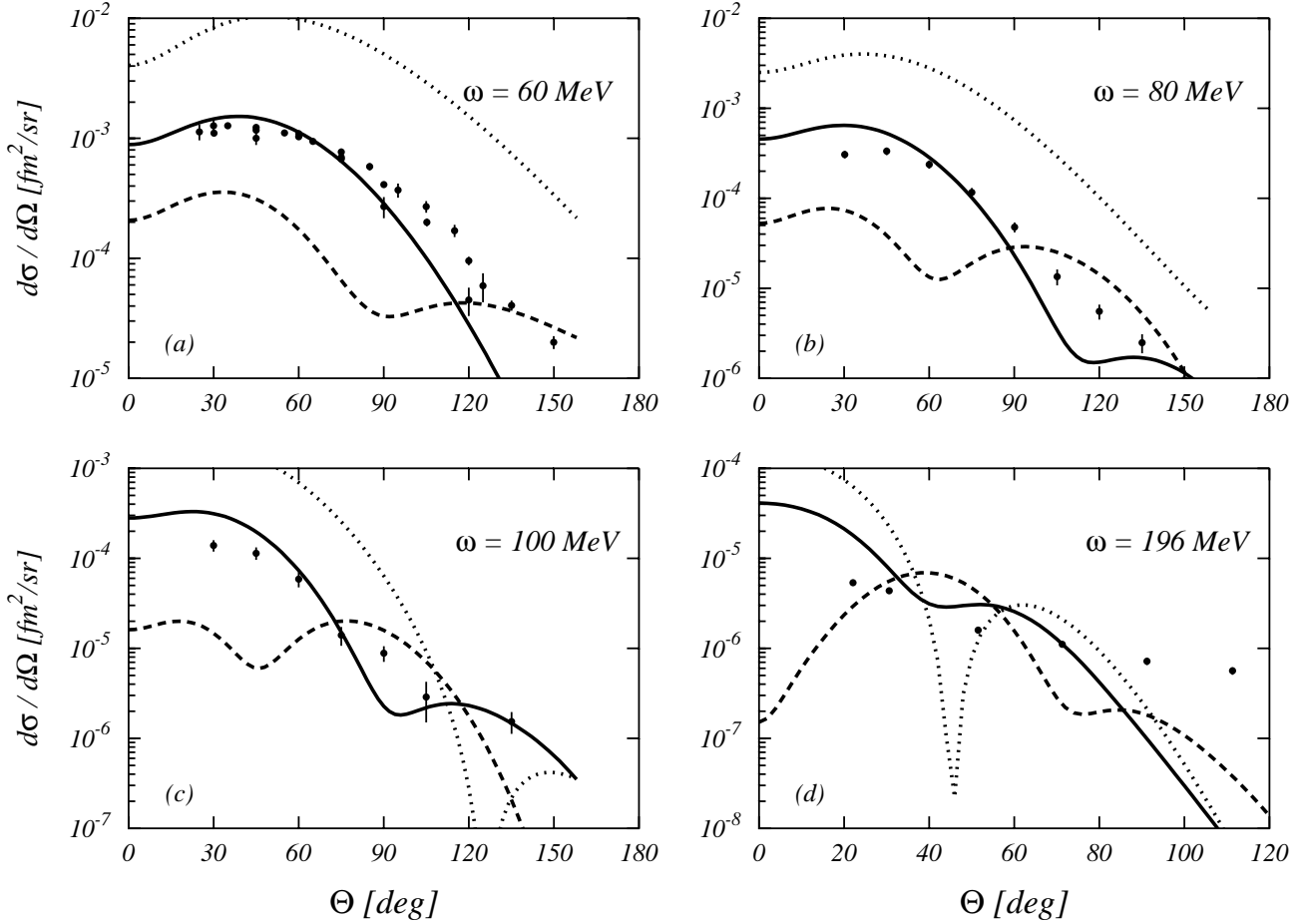}
\vspace{-1.5cm}
\caption{\small Cross sections of the 
  $^{16}$O($\gamma$,p$_0$)$^{15}$N  
  reaction for various energies of the photon. The full
  lines have been obtained using, for the particle states,  the
  energy dependent optical potential  
  of Ref. \protect\cite{sch82}, the dashed
  lines with the real potential
  potential of Ref. \protect\cite{ari96}
  also used for the hole states, and the dotted lines
  have been obtained in PWIA. The 60, 80 and 100 MeV data are those
  already shown in Fig. \protect\ref{fig:gprpa}. The 196 MeV data are
  from Ref. \protect\cite{ada88}.
 }
\label{fig:gpopt}
\end{figure}
%%%%%%%%%%%%%%%%%%%%%%%%%%%%%%%%%%%%%%%%%%%%%%%%%%%%%%%%%%%%%%%%%%%%%

The  angular distributions of the 
$^{16}$O($\gamma$,p$_0$)$^{15}$N cross sections calculated for
various photon energies by using different potentials
are compared in Fig. \ref{fig:gpopt} with the experimental data 
of Refs. \cite{fin77}-\cite{mil95,ada88}.
The agreement between data and the cross section
obtained by using the optical potential is remarkable. 
Despite the differences
in the angular distributions, the order of magnitude of the cross
sections is correct, contrary to the results obtained with the real mean
field (dashed curves), always smaller than the data, 
and those obtained in PWIA (dotted curves), always larger than the data.

We should mention at this point that the RPA calculations of
Refs. \cite{cav84,ryc88}, done with Skyrme type interactions,
reproduce both proton and neutron emission data better than our RPA
results. This indicates that in some parameterization of this
interaction, the FSI effects are effectively considered. 
However the results are extremely sensitive to the choice of the
parameters, as it is shown in Refs. \cite{cav84,ryc88} where results
obtained with different interactions are compared.

Henceforth we shall investigate the proton emission cross sections by
using the direct knock-out model (no RPA) with the 
energy dependent optical potential of Schwandt {\it et al.} \cite{sch82}. 

\subsection{The electromagnetic currents}
\label{mec}

The effects of the SRC we aim to disentangle, are certainly small in 
comparison with the total size of the cross section.  It is therefore
mandatory to control all the approximations of our model, to be sure
that the size of the effects we found do not fall within the
uncertainty of our theoretical hypotheses. In this subsection we
discuss those hypotheses related to the treatment of the
electromagnetic currents.

The usual treatment of the photonuclear processes is based on the Long
Wave Approximation (LWA) and on the Siegert's theorem which allows one to
substitute the convection current with the charge density operator
\cite{bla52}. We have investigated the validity of these
approximations in the energy region above the giant resonance and our
conclusions are analogous to those obtained in
Refs. \cite{ryc87,ryc88}. 
The two main points are that, in this region, the LWA
starts to lose its validity and that multipole
excitations other than the 1$^-$ become important.
For these reasons in our calculations we used the
explicit expression of the convection current operator and in the
evaluation of the ($\gamma$,p) cross section we inserted all the
electric and magnetic multipoles up to J=12 (see Eq. (\ref{psif})).
This last requirement was necessary to ensure the numerical stability
at large values of the emission angle, even at the highest photon 
energies investigated.

\begin{figure}
\vspace*{-3cm}
\hspace*{.5cm}
\includegraphics[bb=50 50 500 680,angle=90,scale=0.8]
{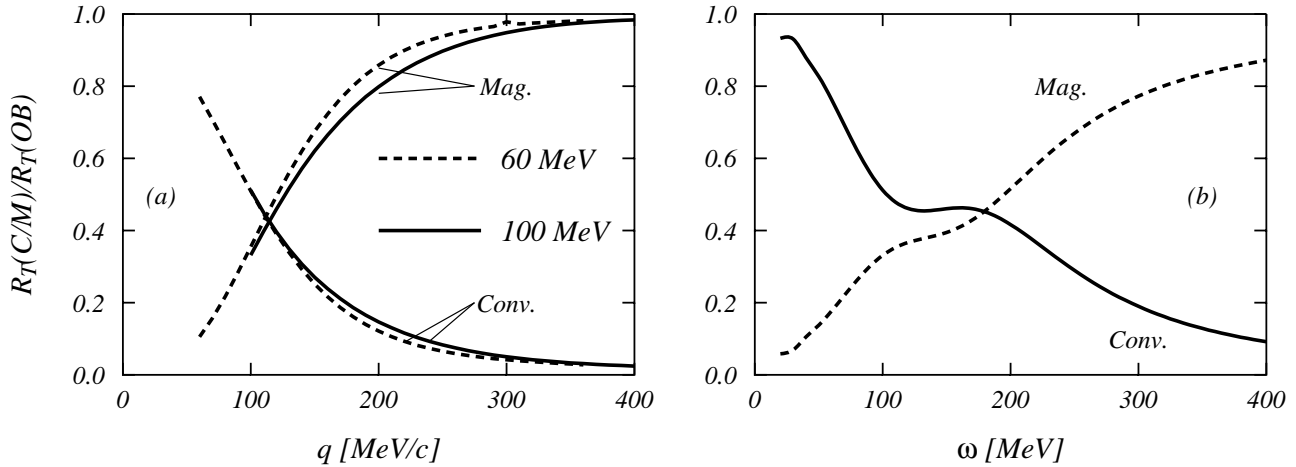}
\vspace{-4.5cm}
\caption{\small Ratio between transverse responses calculated  by
  using only
  convection or magnetization current and the full OB
  current response. In the panel $(a)$ the excitation energies have
  been fixed and the momentum transfer $q$ has changed. In the panel
  $(b)$ the ratios have been calculated at the photon point for
  various values of the excitation energy.
 }
\label{fig:convmag}
\end{figure}

We have studied the relevance of the OB magnetization current, Eq.
(\ref{mag}), usually neglected in the study of photo-nuclear
reactions \cite{gia85,bla52,sar93,ryc88}.  
The ratios between the transverse responses, Eq. (\ref{rt}),
calculated using only the convection or the magnetization current and
the responses obtained with the full OB current are shown in Fig.
\ref{fig:convmag}.

%%%%%%%%%%%%%%%%%%%%%%%%%%%%%%%%%%%%%%%%%%%%%%%%%%%%%%%%%%%%%%
%%%%%%%%%%%%%%%%%%%%%%%%%%%%%%%%%%%%%%%%%%%%%%%%%%%%%%%%%%%%%%%%%%

The panel $(a)$ of the
figure show that at the photon point ($q=\omega$) the contribution
of the convection current is larger than that of the magnetization
current.  With the increase of the momentum transfer, the
relative importance of the two currents is interchanged. In the
quasi-elastic regime the magnetic current dominates as it is well
known \cite{ama01}.
Even at the photon point the relative importance of the magnetization
current increases with increasing energy.  This trend is shown
in the panel $(b)$ of the figure, where the relative ratios
calculated at the photon points are given as a function of the
excitation energy.

The effects of the MEC on the photonuclear reactions are estimated to
be more important than in electron scattering.  This information comes
from indirect evaluation of MEC in RPA calculations based on the
charge-current continuity equation \cite{sar93,ryc88,gar81} and also
from their explicit calculation as it is done in Refs.
\cite{ryc92}-\cite{gai00}.

%%%%%%%%%%%%%%%%%%%%%%%%%%%%%%%%%%%%%%%%%%%%%%%%%%%%%%%%%
\begin{figure}
\vspace*{-3.5cm}
\hspace*{.5cm}
\includegraphics[bb=50 50 500 680,angle=90,scale=0.8]
{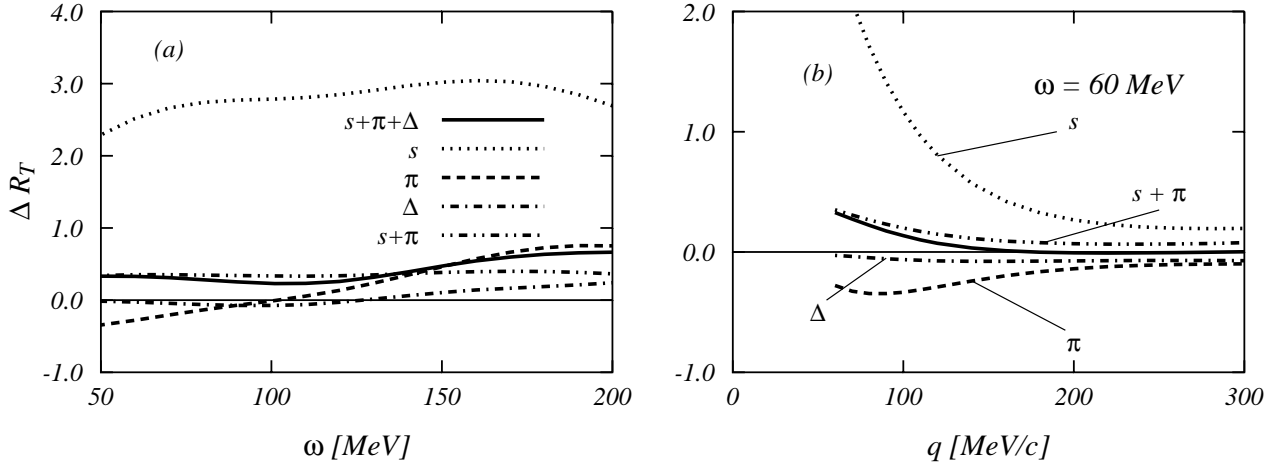}
\vspace{-4.5cm}
\caption{\small 
 Normalized differences, Eq. (\protect\ref{ndiff}), obtained by adding
 to the OB currents the various MEC terms indicated in the figure.
 The normalized differences have been calculated at the photon point 
 (panel $(a)$) and at fixed excitation energy $\omega =60$~MeV as a 
 function of the momentum transfer $q$ (panel $(b)$).    
 The labels indicates the MEC
 included: (s) the seagull, ($\pi$) the pionic, ($\Delta$) the
 $\Delta$ current, and (s + $\pi$) both seagull and pionic
 currents. The full lines have been obtained by considering
 all the currents.
 }
\label{fig:qwmec}
\end{figure}
%%%%%%%%%%%%%%%%%%%%%%%%%%%%%%%%%%%%%%%%%%%%%%%%%%%%%%%

In order to discuss the effects of the MEC we show
in Fig. \ref{fig:qwmec} normalized differences defined as 
\beq
\Delta R_T(q,\omega)= \frac { R_T(OB+MEC) - R_T(OB) } { R_T(OB) } \, ,
\label{ndiff}
\eeq
where we have indicated with $OB + MEC$ the responses obtained by 
adding the various MEC terms to the OB currents. 
In the panel $(a)$ of the figure the normalized differences calculated
at the photon point are shown as a function of the excitation
energy. In the panel $(b)$ as a function of the momentum transfer for
fixed excitation energy $\omega =60$~MeV.

The results presented in Fig. \ref{fig:qwmec} shows that 
the inclusion of the seagull term only, as it is done for example in
Refs. \cite{ben94,gai00}, largely overestimates the role of the MEC.
The pionic current alone reduces the OB response at energies smaller than
100 MeV, and it slightly increases it at larger values.  
The presence of 
destructive interference
between seagull and pionic currents is well known in the
studies of the quasi-elastic peak \cite{ama93,ord81}.
The dashed doubly dotted line of the figure shows the result obtained
by including both currents.

The contribution of the $\Delta$ current is relatively
small. We should remark that we consider only the virtual excitation
of $\Delta$, and do not include the pion production channel which
implies the real delta excitation. 
This channel is open above $\omega$=140 MeV.

Panel $(b)$ show that
at the photon point the contribution of the 
seagull term is relatively larger than at higher $q$ values.
The pionic term reduces the OB response for all $q$ values.
The interference between these two terms is the
dominant contribution since the $\Delta$ current is negligible. 
The full line shows that the
inclusion of all the MEC diagrams produces relatively small variations
on the OB currents.  It is worth to notice that the relative difference
becomes smaller with increasing momentum transfer. The photon point is
the place where the MEC effects are largest, on a relative scale.

%%%%%%%%%%%%%%%%%%%%%%%%%%%%%%%%%%%%%%%%%%%%%%%%%%%%%%%%%%%%%%%%%%%%%
\begin{figure}
\vspace*{-3.cm}
\hspace*{.5cm}
\includegraphics[bb=50 50 500 680,angle=90,scale=0.8]
{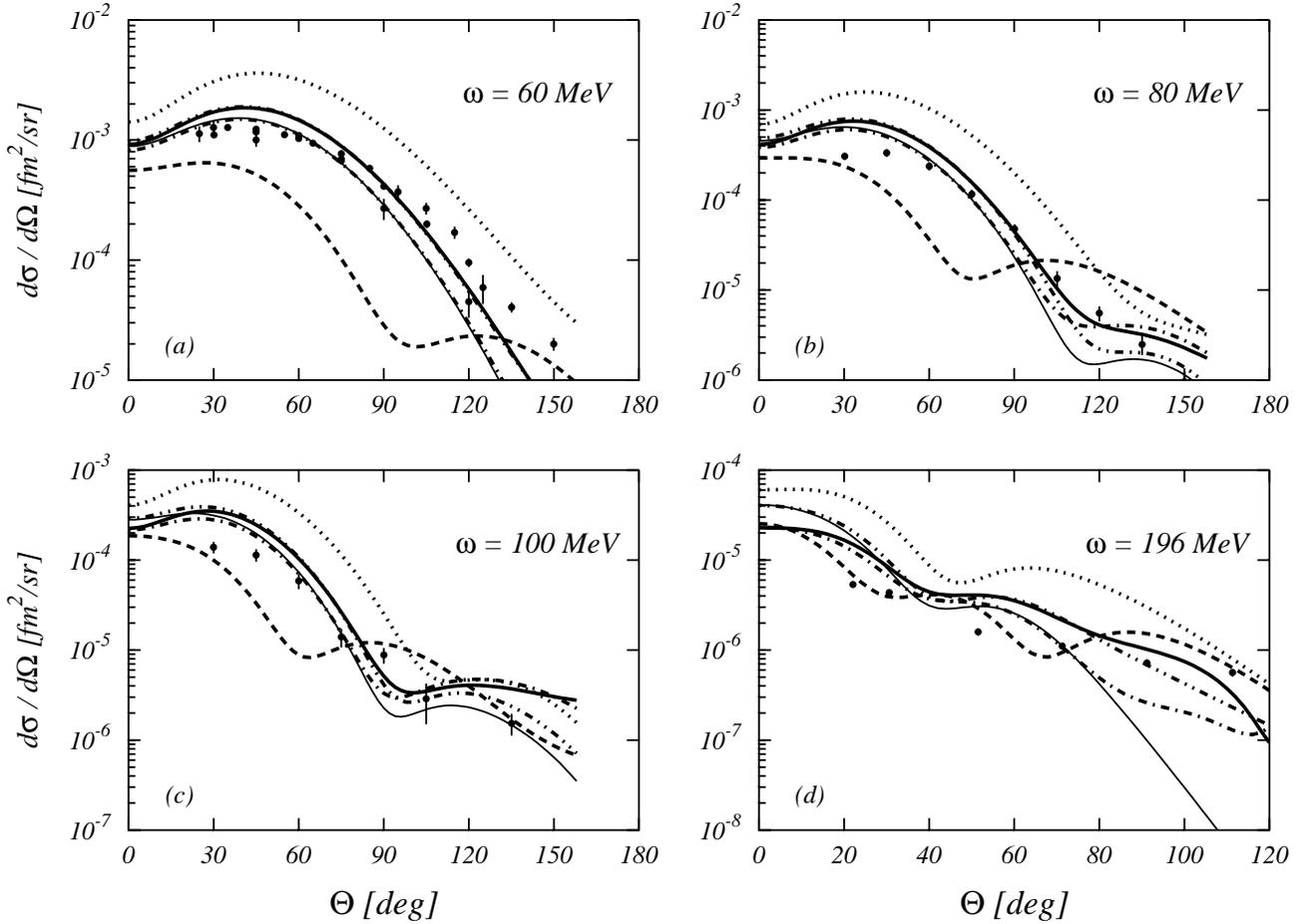}
\vspace{-1.5cm}
\caption{\small Angular distribution of the  
 $^{16}$O($\gamma$,p$_0$)$^{15}$N 
 cross section for various energies of the photon. The thin
 full lines show the OB results, the other lines have been obtained by
 adding the various terms of the MEC: dotted lines, OB+seagull, dashed
 lines, OB+pionic, dashed-dotted, OB+$\Delta$, dashed-doubly-dotted,
 OB+seagull+pionic. The thick full lines show the results obtained
 once all the currents are included. Data as in
  Fig. \protect\ref{fig:gpopt}.
 }
\label{fig:gpmec}
\end{figure}
%%%%%%%%%%%%%%%%%%%%%%%%%%%%%%%%%%%%%%%%%%%%%%%%%%%%%%%%%%%%%%%%%%%%

In Fig. \ref{fig:gpmec} we compare the angular distributions of the 
$^{16}$O($\gamma$,p$_0$)$^{15}$N cross sections calculated for
various photon energies. The calculations have been done with
the optical potential and the thin full lines represent the results
obtained with the OB currents only. The other lines have been obtained
by adding the various
MEC terms. 
The
contribution of the seagull current (dotted lines) noticeably
increases the cross section, but the effect of the pionic current
(dashed lines) goes in the opposite direction. When both currents are
considered (dashed-doubly dotted lines), the results are only slightly
above the OB ones. The excitation of the virtual $\Delta$
(dashed-dotted lines) is not important but at high values of $\Theta$
and at high photon energies.
The total results (thick full lines) are slightly above the OB ones,
the differences become larger at high values of $\Theta$.

The results
obtained with the OB currents only (thin full lines) steeply decrease
with increasing $\Theta$. At high $\Theta$ values the inclusion of
seagull and pionic currents pushes up the cross section by an order of
magnitude.

\subsection{The short-range correlations}
\label{src}
\begin{figure}
\vspace*{-3.3cm}
\hspace*{.5cm}
\includegraphics[bb=50 50 500 680,angle=90,scale=0.8]
{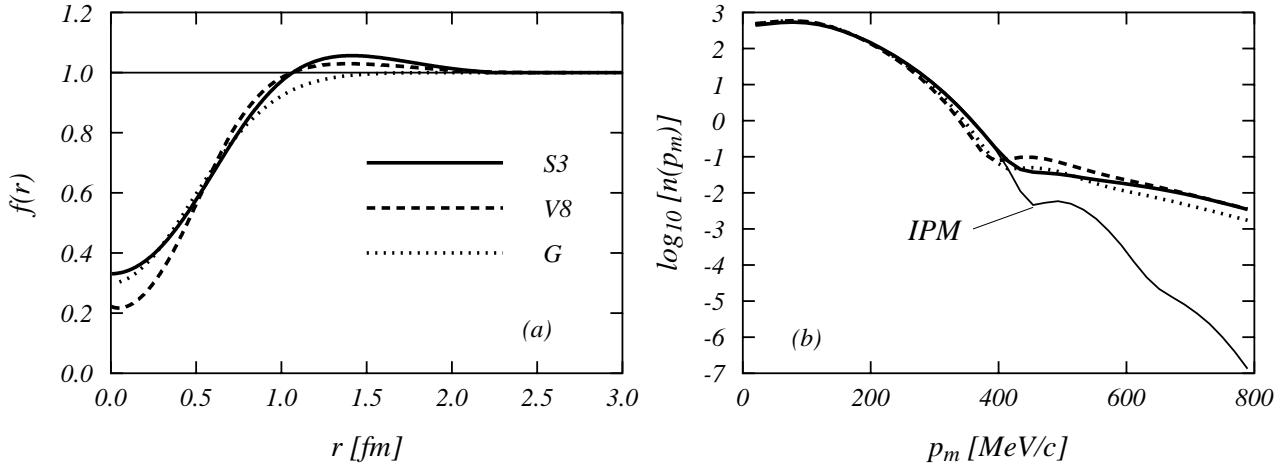}
\vspace{-4.5cm}
\caption{\small Left panel: correlation functions considered in our
  calculations. Right panels: momentum distributions calculated with
  the correlations of the left panel.
  }
\label{fig:corr}
\end{figure}

We have investigated the effects of the SRC by evaluating the
$(\gamma,p)$ cross sections with different correlation functions.  
The hole wave functions and the correlation
functions have been fixed in Ref. \cite{ari96} to minimize the nuclear
hamiltonian calculated with the S3 interaction of Afnan and Tang
\cite{afn68}. We used the two types of correlation
function selected in Ref. \cite{ari96}: a first one obtained with the
Euler procedure, labelled S3 in the panel $(a)$ of Fig.
\ref{fig:corr}, and a second one of gaussian type, labelled G.  For
sake of comparison we also used the scalar part of the state dependent
correlation function taken from \cite{fab00} where the hamiltonian
expectation value has been calculated with a more realistic
interaction: the V8' Argonne interaction \cite{pud97}.

%%%%%%%%%%%%%%%%%%%%%%%%%%%%%%%%%%%%%%%%%%%%%%%%%%%%%%%%%%%%%%%%%%5
%%%%%%%%%%%%%%%%%%%%%%%%%%%%%%%%%%%%%%%%%%%%%%%%%%%%%%%%%%%%%%%%%%%55

In the panel $(b)$ of Fig. \ref{fig:corr} the ground state momentum
distributions calculated with the first order model of Ref.
\cite{ari97} are shown for the three different correlations. The
behavior of the various results show the well known increase at high
momentum values with respect to the IPM result 
\cite{ant88}-\cite{fab01}. 
There are not striking differences between the momentum distributions
obtained with the various correlation functions.

%%%%%%%%%%%%%%%%%%%%%%%%%%%%%%%%%%%%%%%%%%%%%%%%%%%%%%%%%%%%%%%%%%%
\begin{figure}
%\vspace*{-3.cm}
\hspace*{.3cm}
\includegraphics[bb=10 50 500 680,angle=0,scale=0.8]
{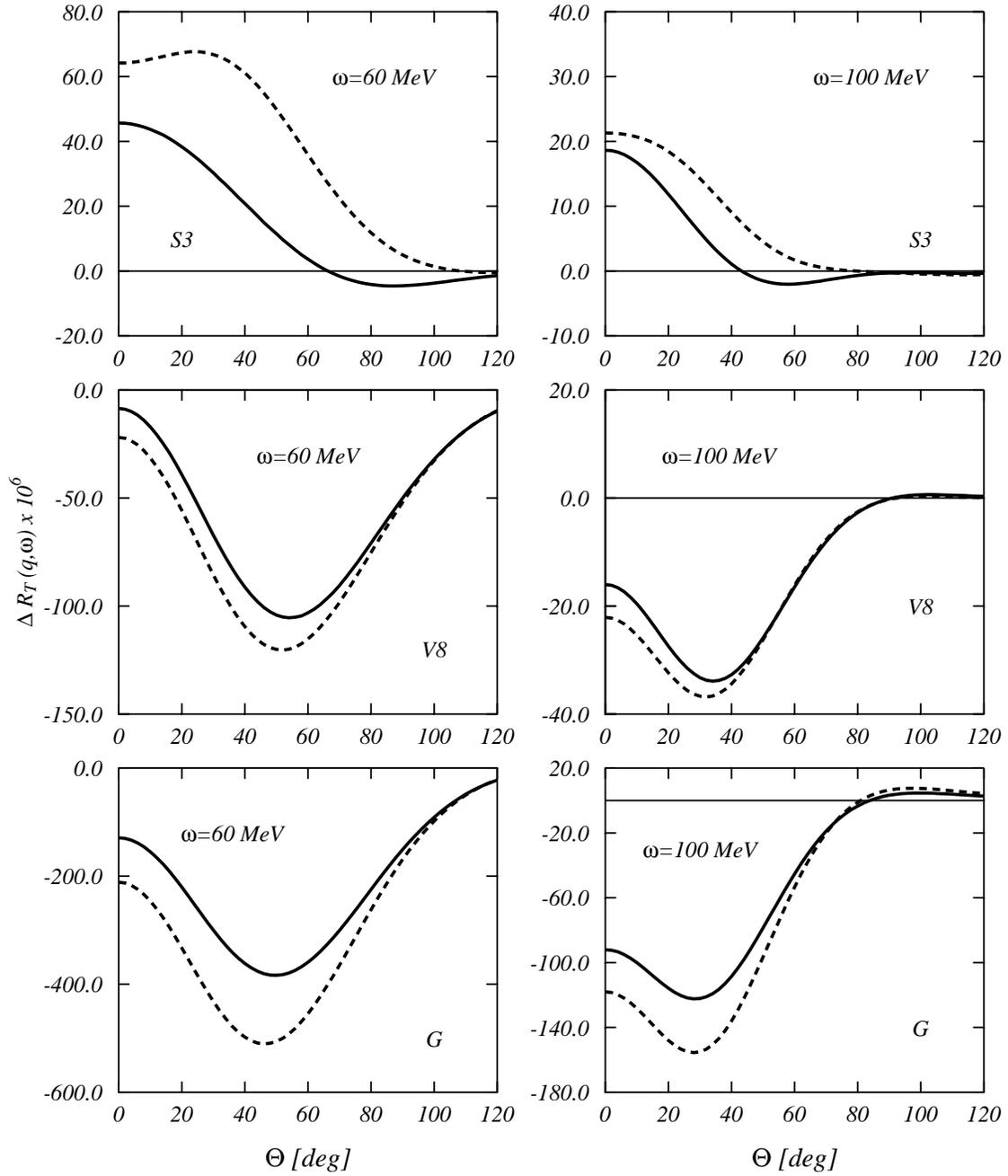}
\vspace{-1.5cm}
\caption{\small Normalized difference between correlated and uncorrelated
  responses for photon energies of 60 (left panels) and 
  100 (right panels) MeV. The dashed lines have been calculated with 2 point
  diagrams only, the full lines with both 2 and 3 point diagrams.
  The upper, medium and lower panels correspond to calculations performed
  with the S3, V8 and gaussian correlations, respectively.
  }
\label{fig:cor23}
\end{figure}
%%%%%%%%%%%%%%%%%%%%%%%%%%%%%%%%%%%%%%%%%%%%%%%%%%%%%%%%%%%%%%%%%%%55

Since the novelty of our approach is the inclusion of the three point
diagrams we have studied the importance of these terms.  In Fig.
\ref{fig:cor23} we show the normalized difference of Eq. (\ref{ndiff})
evaluated by adding to the OB terms the correlated 
two- and three-point diagrams instead of the MEC.
The full lines have been obtained by considering all the diagrams of
Fig. \ref{fig:mey} and the dashed lines by adding to the IPM response
the two-point diagrams only.

These results agree with the findings of Refs. \cite{co01} and
\cite{mok01} in the (e,e') and (e,e'p) reactions.  
The effect of the two-point diagrams, whatever it is, is lowered by the
inclusion of the three point terms.
It is interesting to notice that the effect of the S3
correlation is smaller than that of the other two correlations
and it has opposite sign.

%%%%%%%%%%%%%%%%%%%%%%%%%%%%%%%%%%%%%%%%%%%%%%%%%%%%%%%%%555
\begin{figure}
\hspace*{.5cm}
\includegraphics[bb=50 50 500 680,angle=90,scale=0.8]
{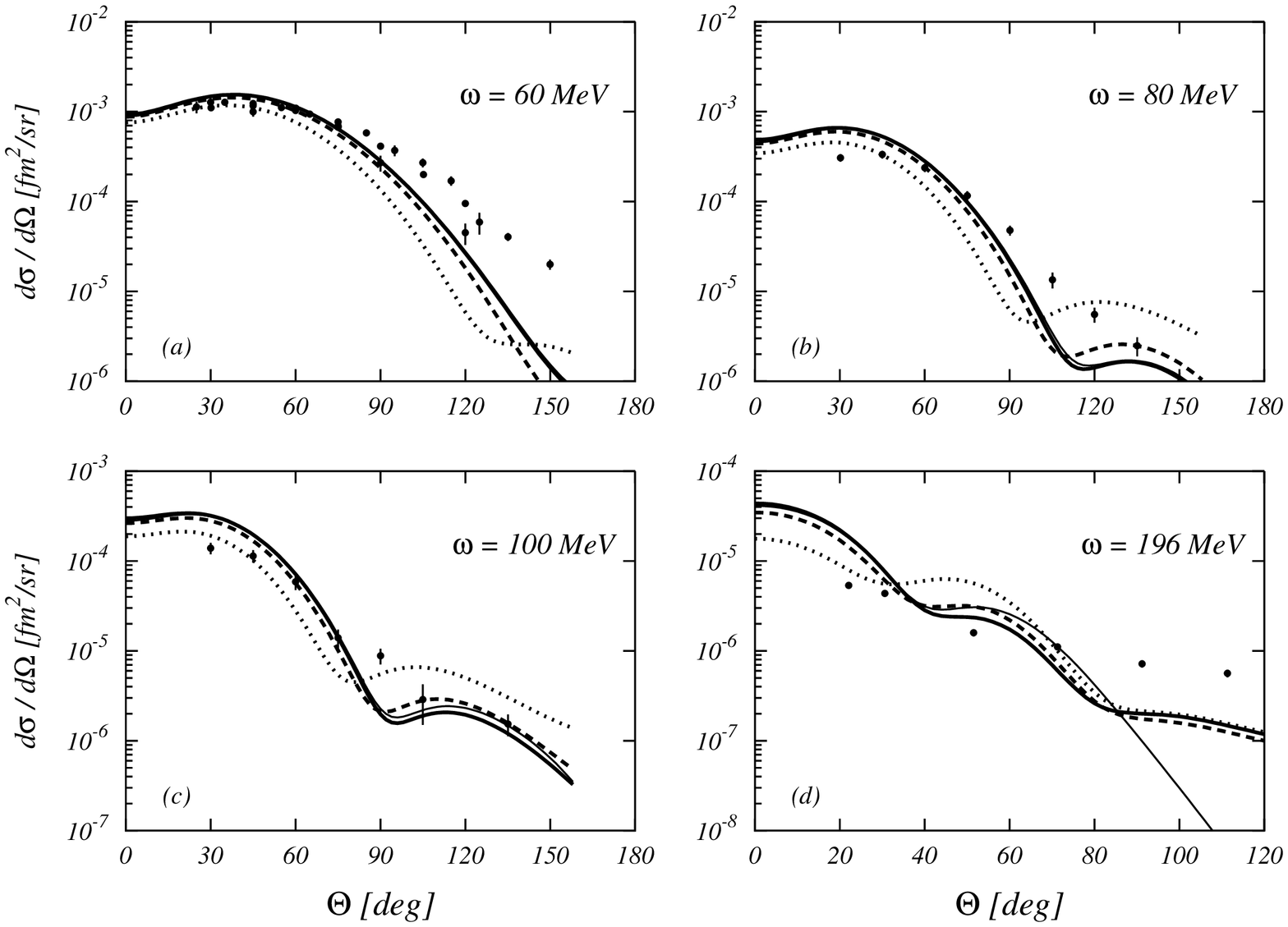}
\vspace{-1.5cm}
\caption{\small Angular distributions calculated with the inclusion of
  the SRC. 
  The thin full lines are the IPM results, the thick full lines 
  show the results obtained with the S3 correlation,
  the dashed lines with the V8 correlation function and the dotted
  lines with the gaussian correlation function.
 }
\label{fig:gpcor}
\end{figure}
%&&&&&%%%%%%%%%%%%%%%%%%%%%%%%%%%%%%%%%%%%%%%%%%%%%%5555

The angular distributions of the $^{16}$O($\gamma$,p$_0$)$^{15}$N
cross sections calculated for different photon energies are shown in
Fig. \ref{fig:gpcor}.  The thin full lines
show the pure IPM results. The thicker full, dashed and dotted lines
have been obtained by including the contribution of the S3, V8
and gaussian correlations respectively.
Also in this case the three correlations produce rather different
effects. 

%%%%%%%%%%%%%%%%%%%%%%%%%%%%%%%%%%%%%%%%%%%%%%%%%%%%%%%%%%%%%%%%%%%%%
\begin{figure}
\includegraphics[bb=10 50 500 680,angle=0,scale=0.8]
{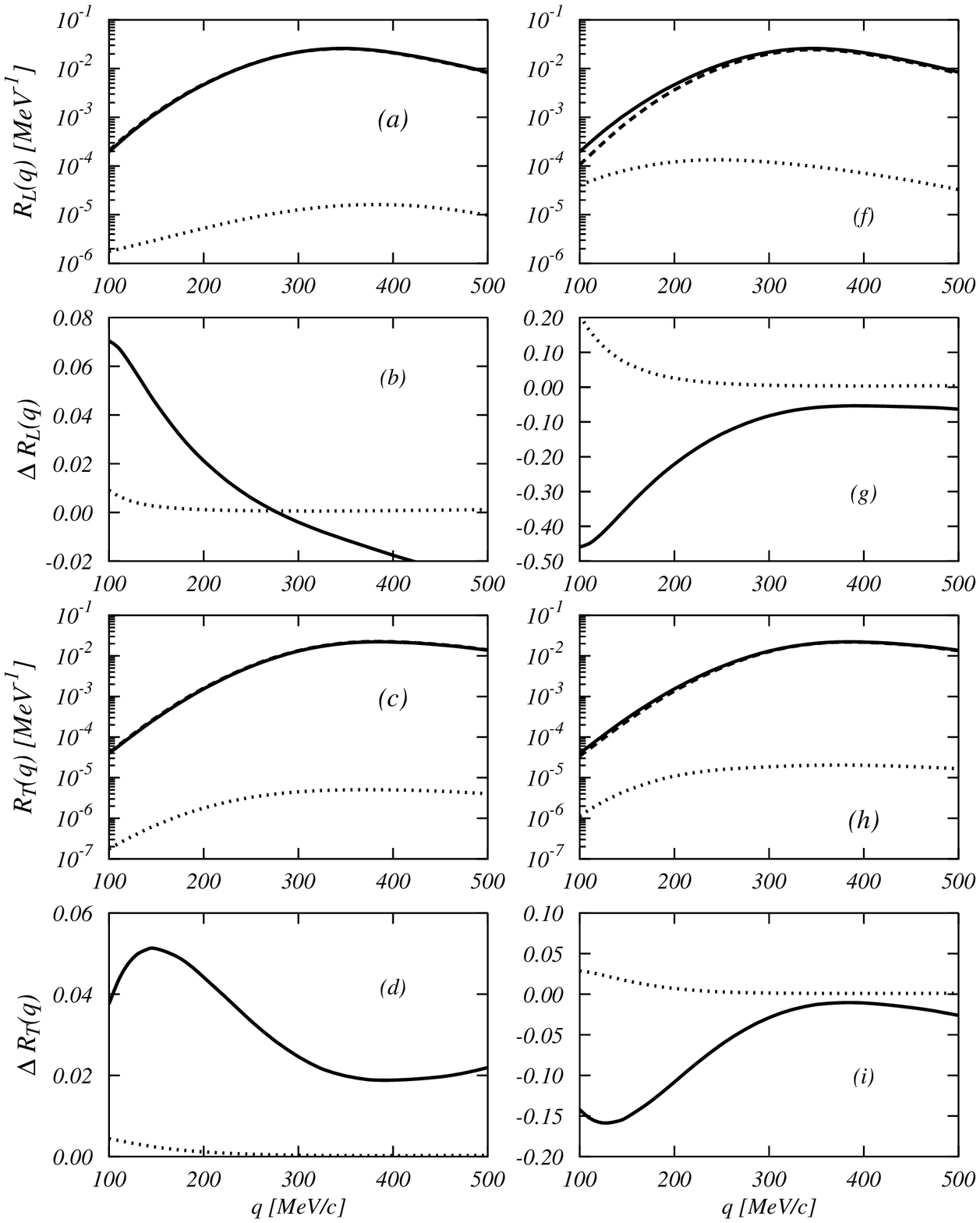}
\vspace{-1.cm}
\caption{\small Inclusive responses and normalized ratios (see text)
  as a function of the momentum transfer at $\omega=100$ MeV.
  Left panels refer to the S3 correlation; right ones to the
  gaussian correlations. The $(a)$, $(b)$, $(f)$ and $(g)$ panels 
  correspond to the
  longitudinal response (not present in photonuclear processes) and
  the other ones to the transverse response. The full lines of 
  panels $(a)$, $(c)$, $(f)$ and $(h)$ are the OB responses, the dashed 
  ones the
  OB+correlations responses and the dotted lines the responses
  obtained only with the correlation terms. The full lines of 
  panels $(b)$, $(d)$, $(g)$ and $(i)$ show the difference between OB and
  OB+correlation responses normalized to the OB response. The
  dotted lines the ratio between the responses
  obtained only with the correlation terms and the OB responses.
  }
\label{fig:qcor}
\end{figure}
%%%%%%%%%%%%%%%%%%%%%%%%%%%%%%%%%%%%%%%%%%%%%%%%%%%%%%%%%%%%%%%%%%%%%%%5

The sensitivity of the results to the correlation is
larger than in the case of electron scattering. We have investigated
the source of this relatively high sensitivity by calculating the
(e,e') electromagnetic responses at fixed excitation energy
($\omega$=100 MeV) for various values of the momentum transfer. 
The results of these calculations are shown in Fig. \ref{fig:qcor}.
All the results shown in the left panels have been obtained by using
the S3 correlation while those of the right panels with the gaussian
correlation. The $(a)$ and $(f)$ panels show the longitudinal responses,
not present in reactions with real photons. The full lines are the IPM
responses and the dashed lines have been obtained by adding the
correlations. Referring to Fig. \ref{fig:mey}, the full lines have been
obtained using only the diagram 1.1 and the dashed lines with all the
diagrams of the figure. In addition to these lines we show the
responses calculated without the OB terms. This means that we used all
the diagrams labelled 2 and 3 in Fig. \ref{fig:mey} and we eliminated
the contribution of the 1.1 diagram. The results of these calculations
are shown by the dotted lines. The analogous results for the
transverse responses are shown in the panels $(c)$ and $(h)$ of the figure. 
To emphasize the difference between the various responses,
in the panels $(b)$, $(d)$, $(g)$ and $(i)$ we show the normalized 
differences.

As expected, for all the momentum transfer values calculated, the OB
responses are orders of magnitude larger than the responses obtained
only with the correlation terms. Also the qualitative behavior of
the responses as a function of $q$ is quite different. While the OB
responses increase by more than two order of magnitude with increasing
$q$ the change of the correlated responses is much more limited. The
consequence of this fact is that, relatively speaking,
at the photon point IPM and
correlated responses are closer in magnitude than at higher $q$.  For
this reason the photon cross sections are relatively more sensitive to
the SRC than the electron scattering processes.
The comparison between the normalized differences, shown in the
$(b)$, $(d)$, $(g)$ and $(i)$ panels, is a measure of the importance of the
interference effects between OB and correlation transition amplitudes.
As expected the most important part of the correlation effects is due
to these interference terms.

The observation of the angular distributions of Fig. \ref{fig:gpcor}
shows that at high values of the photon energy the
correlation effects show up at large scattering angle. The IPM cross
sections are almost overlapped to the thick full lines in Fig.
\ref{fig:gpcor}. At 196 MeV the IPM cross
sections show a fast decrease for large nucleon emission angles while
the correlated cross sections indicates a less rapid fall. This
behavior remember that of the momentum distribution of Fig.
\ref{fig:corr}. In effect, by transforming the value of the emission
angle $\Theta$ in the value of the momentum of the target nucleon
$p_i$, it is possible to verify that
the values of $p_i$ probed go from 400 to 700 MeV/c. 
This is the
interval where the correlated nucleon momentum distributions start to
differ from that obtained with the IPM.

\subsection{Comparison with data}
\label{data}
\begin{figure}
\hspace*{.5cm}
\includegraphics[bb=50 50 500 680,angle=90,scale=0.8]
{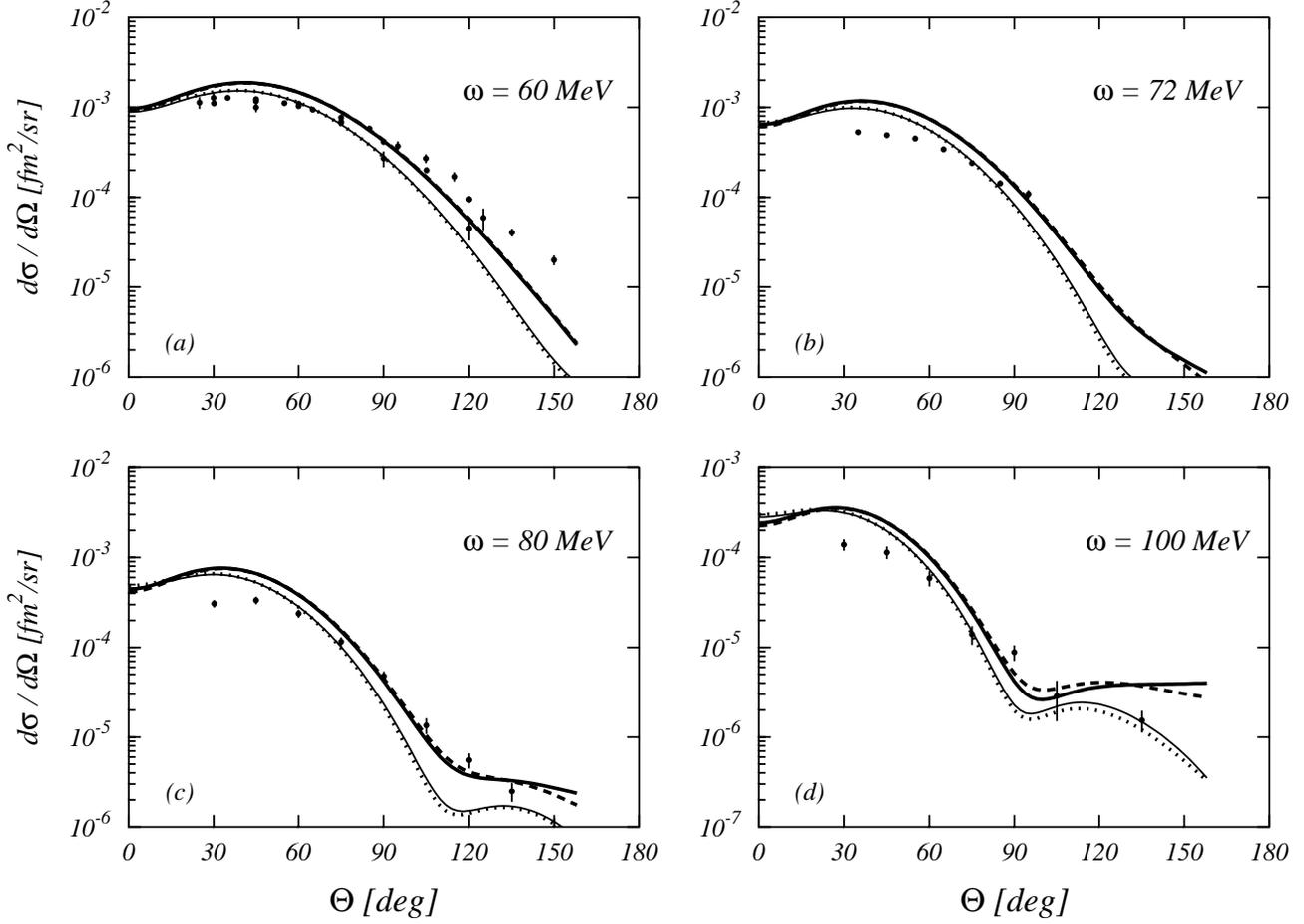}
\vspace{-1.5cm}
\caption{\small Angular distributions of the  
  $^{16}$O($\gamma$,p$_0$)$^{15}$N cross section. The thin full lines have
 been calculated in the IPM by using OB currents only. 
 The dotted lines include the effects of the S3
 correlation. The dashed lines the MEC and the thick full lines all
 the effects. Data are from Refs. \protect\cite{fin77,des93,mil95}.
 }
\label{fig:gptot}
\end{figure}

\begin{figure}
\hspace*{.5cm}
\includegraphics[bb=50 50 500 680,angle=90,scale=0.8]
{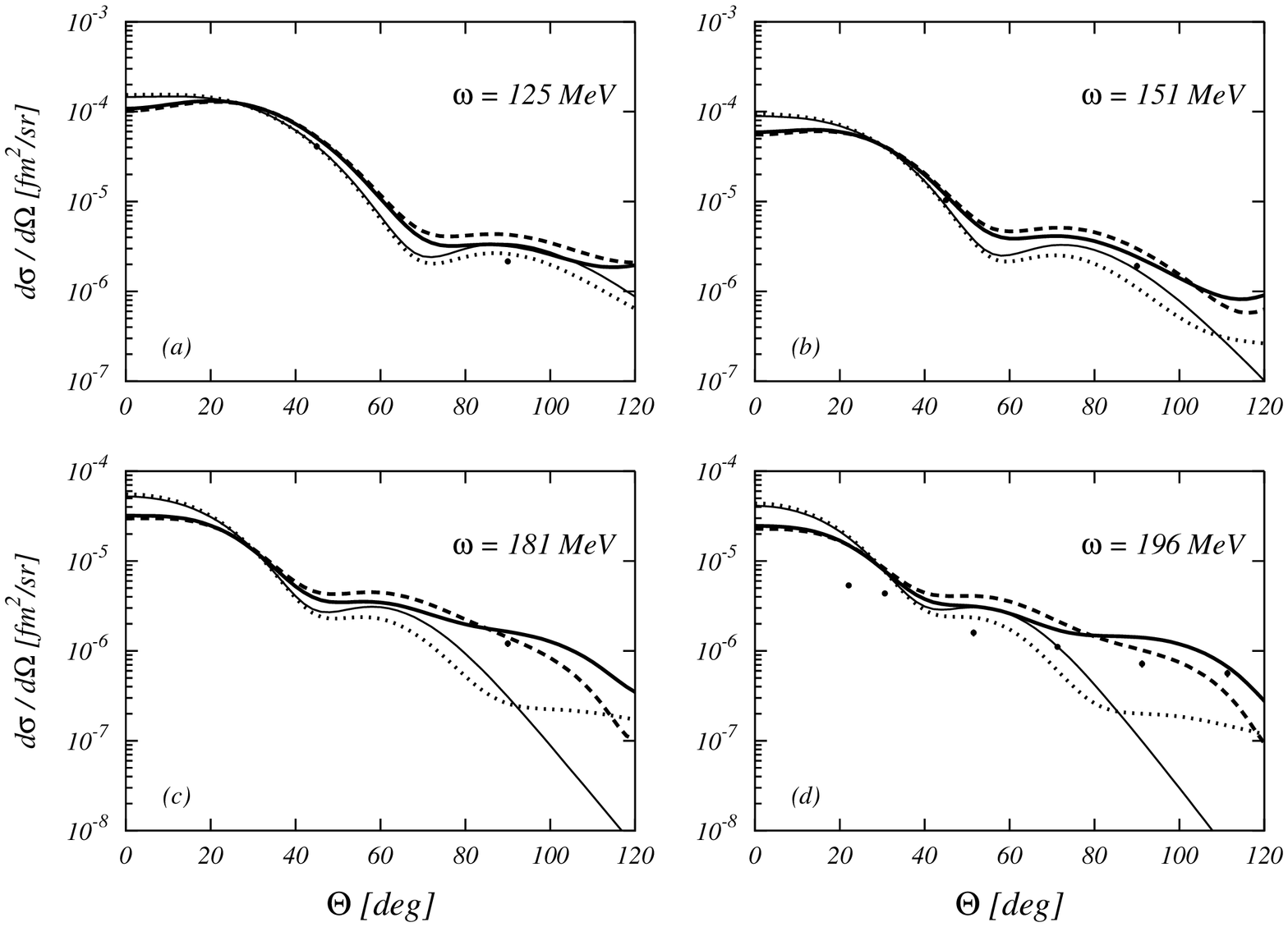}
\vspace{-1.5cm}
\caption{\small The same as in Fig. \protect\ref{fig:gptot} for energies
  above 100 MeV.
  The data at 126, 151 and 180 MeV are from
  Ref. \protect\cite{lei85} and those at 196 MeV from 
  Ref. \protect\cite{ada88}. 
 }
\label{fig:gptot1}
\end{figure}

The results we have so far obtained are summarized in the figures
\ref{fig:gptot} and \ref{fig:gptot1} to discuss their comparison with
the data.
The thin full lines have been calculated within the simple IPM by using
OB currents only. The dotted lines show the
results obtained by adding the S3 SRC obtained with the Euler procedure. 
The inclusion of the MEC produces the dashed
lines. Finally the thick full lines have been obtained by including
both SRC and MEC.
Our calculations overestimate the data at small emission angles. This is
evident in Fig. \ref{fig:gptot} and in Fig. \ref{fig:gptot1}
at $\omega$=196 MeV where more data are available.
The shape of the data at high energy values
is reproduced only because of
the inclusion of SRC and MEC. It is evident that the MEC contributions
are larger than those of the SRC.

%%%%%%%%%%%%%%%%%%%%%%%%%%%%%%%%%%%%%%%%%%%%%%%%%%%%%%%%%%%%%%%%%%%%%%%%%%
%%%%%%%%%%%%%%%%%%%%%%%%%%%%%%%%%%%%%%%%%%%%%%%%%%%%%%%%%%%%%%%%%%%%%%%%%%
%%%%%%%%%%%%%%%%%%%%%%%%%%%%%%%%%%%%%%%%%%%%%%%%%%%%%%%%%%%%%%%%%%%%%%%%%%

\begin{figure}
%\hspace*{.5cm}
\vspace*{-2.5cm}
\includegraphics[bb=50 50 500 730,angle=90,scale=0.8]
{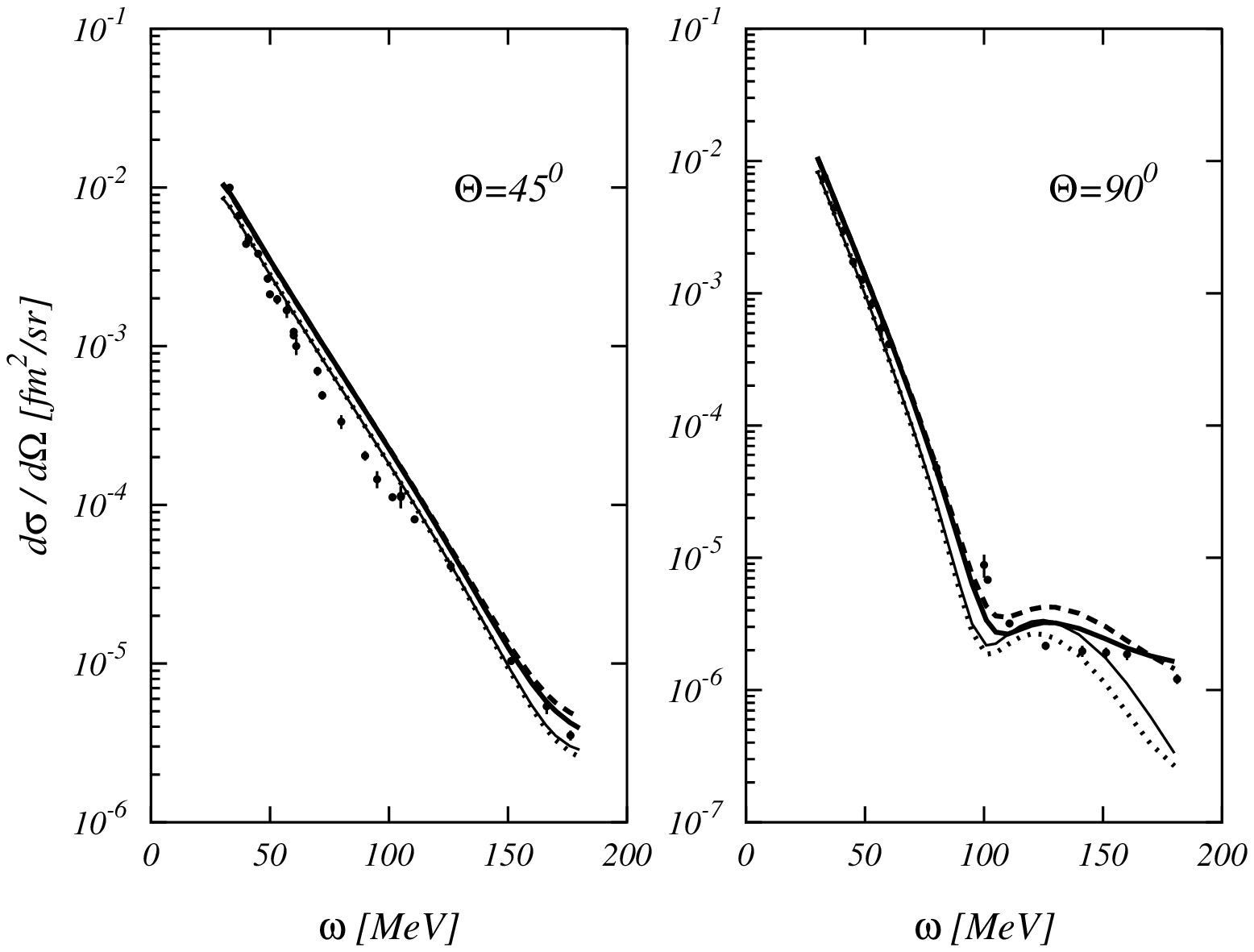}
\vspace{-2.5cm}
\caption{\small Energy dependence of the 
  $^{16}$O($\gamma$,p$_0$)$^{15}$N 
  cross section for various proton emission angles.
  Like in the previous figures, the thin full lines have been obtained
  with the IPM model, the dotted ones by adding the SRC, the dashed ones
  with the MEC and the full thick lines show the total results.  
  Data are from \protect\cite{fin77,des93,mil95,lei85}.
 }
\label{fig:edep}
\end{figure}

The energy dependence of the $^{16}$O($\gamma$,p$_0$)$^{15}$N cross
section for fixed values of the emission angle is shown in Fig.
\ref{fig:edep}. The meaning of the lines is the same as in Fig.
\ref{fig:gptot}. For $\Theta$=45$^0$ our calculations show an almost
perfect exponential decay. On the other hand the data are not so
perfectly aligned. The discrepancy between our calculations and the
data is more evident between 50 and 100 MeV. The shapes of the data
and of our results for $\Theta$=90$^0$ are characterized by two
different trends: an exponential decay up to 100 MeV and an almost
flat behavior at higher energies. This high energy tail is dominated
by the MEC. We also observe that for $\omega >$ 100 MeV the SRC effect
consists in a reduction of the cross section. 

%%%%%%%%%%%%%%%%%%%%%%%%%%%%%%%%%%%%%%%%%%%%%%%%%%%%%%%%%%%%%%%%%%%%%%%%%%
%%%%%%%%%%%%%%%%%%%%%%%%%%%%%%%%%%%%%%%%%%%%%%%%%%%%%%%%%%%%%%%%%%%%%%%%%%
\begin{figure}
\vspace*{-3cm}
\hspace*{.5cm}
\includegraphics[bb=50 50 500 680,angle=90,scale=0.8]
{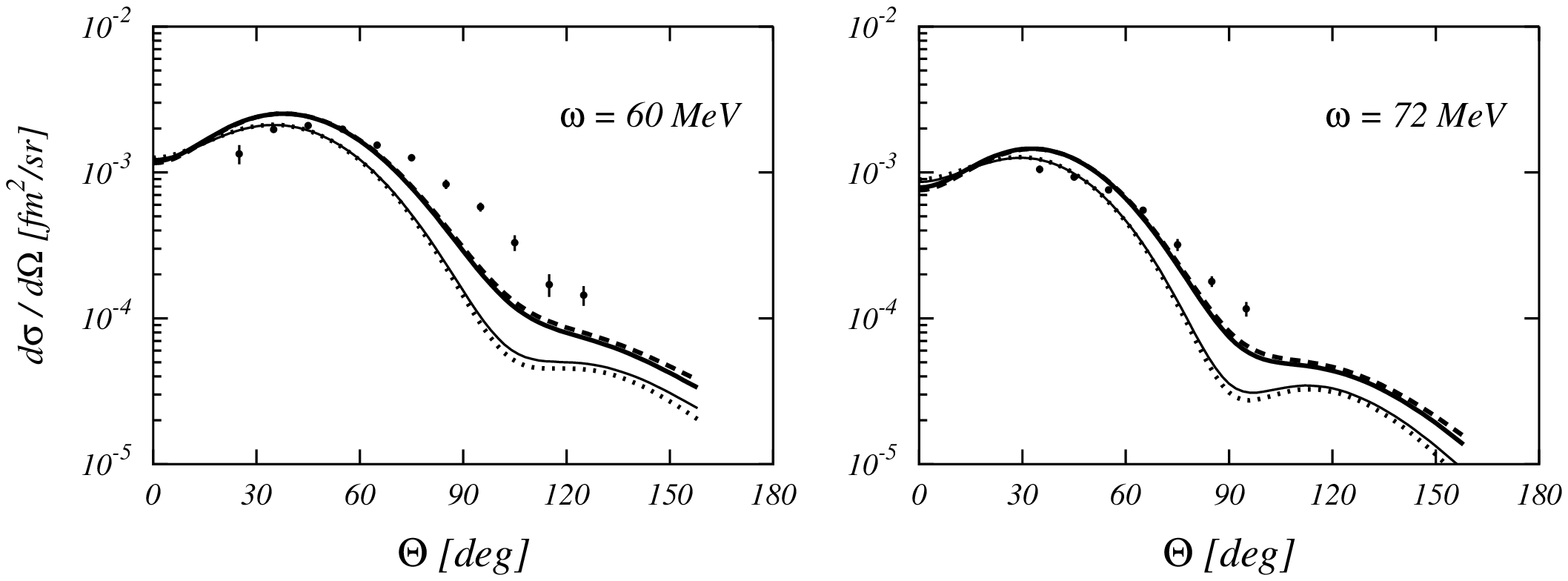}
\vspace{-3.5cm}
\caption{\small Angular distributions for the emission of a proton from
  the 1p$_{3/2}$ level. The meaning of the lines is the same as in
  Fig. \protect\ref{fig:gptot}. Data are from Ref. \protect\cite{mil95}.
 }
\label{fig:gp32}
\end{figure}
%%%%%%%%%%%%%%%%%%%%%%%%%%%%%%%%%%%%%%%%%%%%%%%%%%%%%%%%%%%%%%%%%%%%%%%%%%

In Fig. \ref{fig:gp32} we compare the results of our calculations with
the data relative to the emission of the proton from the 1p$_{3/2}$ level
\cite{mil95}. The meaning of the various lines is the same as in
Fig. \ref{fig:gptot}. Also in this case our calculations slightly
overestimate the data at small emission angles and underestimate them
at larger angles. The SRC reduce the cross section while MEC increase
it. 

\begin{figure}
\vspace{-3.5cm}
\hspace*{.5cm}
\includegraphics[bb=50 50 500 680,angle=90,scale=0.8]
{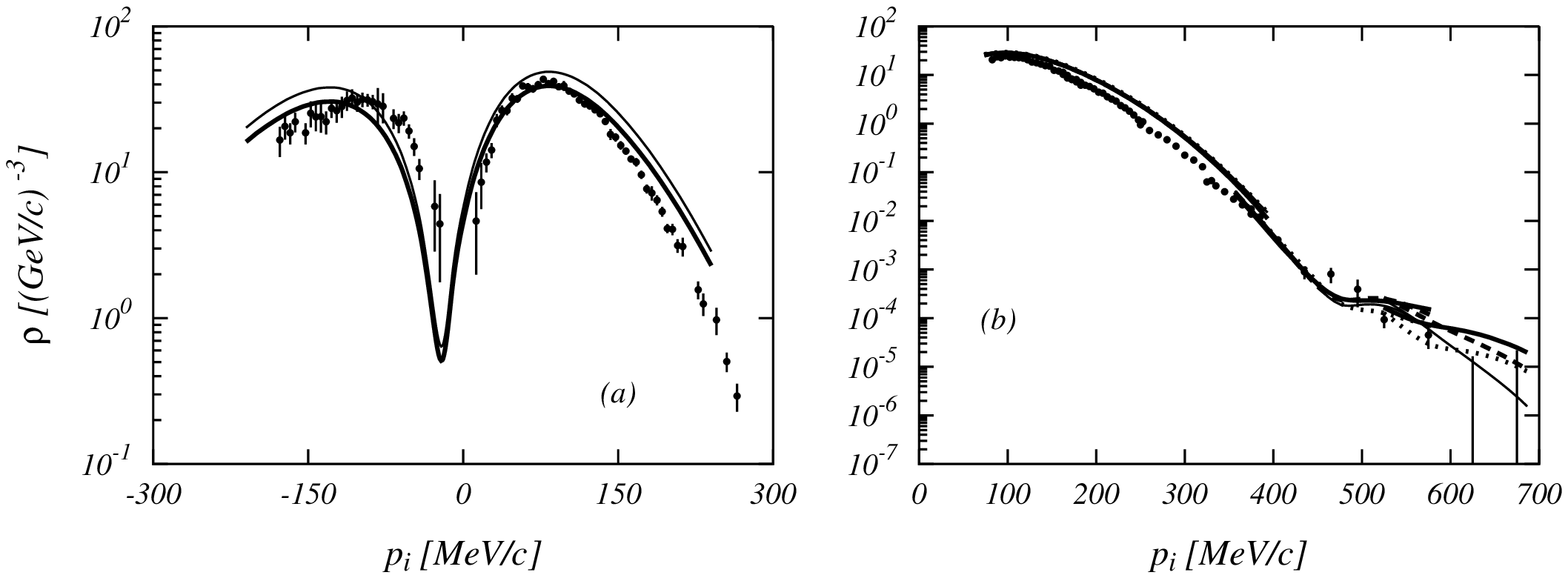}
\vspace{-4cm}
\caption{\small  $^{16}$O(e,e'p$_0$)$^{15}$N cross sections from
  NIKHEF \protect\cite{leu94}, panel $(a)$, and Mainz
  \protect\cite{blo95}, panel $(b)$,
  kinematics. In panel $(a)$ the thin line
  represent the bare calculation while the full line has been
  multiplied by the spectroscopic factor of 0.8. All the curves of the
  panel $(b)$ are multiplied by the spectroscopic factor. The meaning
  of the lines in this panel is the same as in
  Fig. \protect\ref{fig:gptot}. 
  }
\label{fig:eep}
\end{figure}

In all the calculations presented so far we did not make use of
spectroscopic factors, a common practice in the analysis of (e,e'p)
data. We fixed the spectroscopic factor as a reduction factor required
by our $^{16}$O(e,e'p$_0$)$^{15}$N results in order to reproduce the
NIKHEF data of Ref. \cite{leu94}. The comparison between our theory
and the data is shown in the panel $(a)$ of Fig. \ref{fig:eep}. From
this comparison we obtained a spectroscopic factor of 0.8, as in Ref.
\cite{mok01}.

%%%%%%%%%%%%%%%%%%%%%%%%%%%%%%%%%%%%%%%%%%%%%%%%%%%%%%%%%%%%%%%%%%%%%
%%%%%%%%%%%%%%%%%%%%%%%%%%%%%%%%%%%%%%%%%%%%%%%%%%%%%%%%%%%%%%%%%%%%%%
\begin{figure}
\hspace*{.3cm}
\includegraphics[bb=10 50 500 680,angle=0,scale=0.8]
{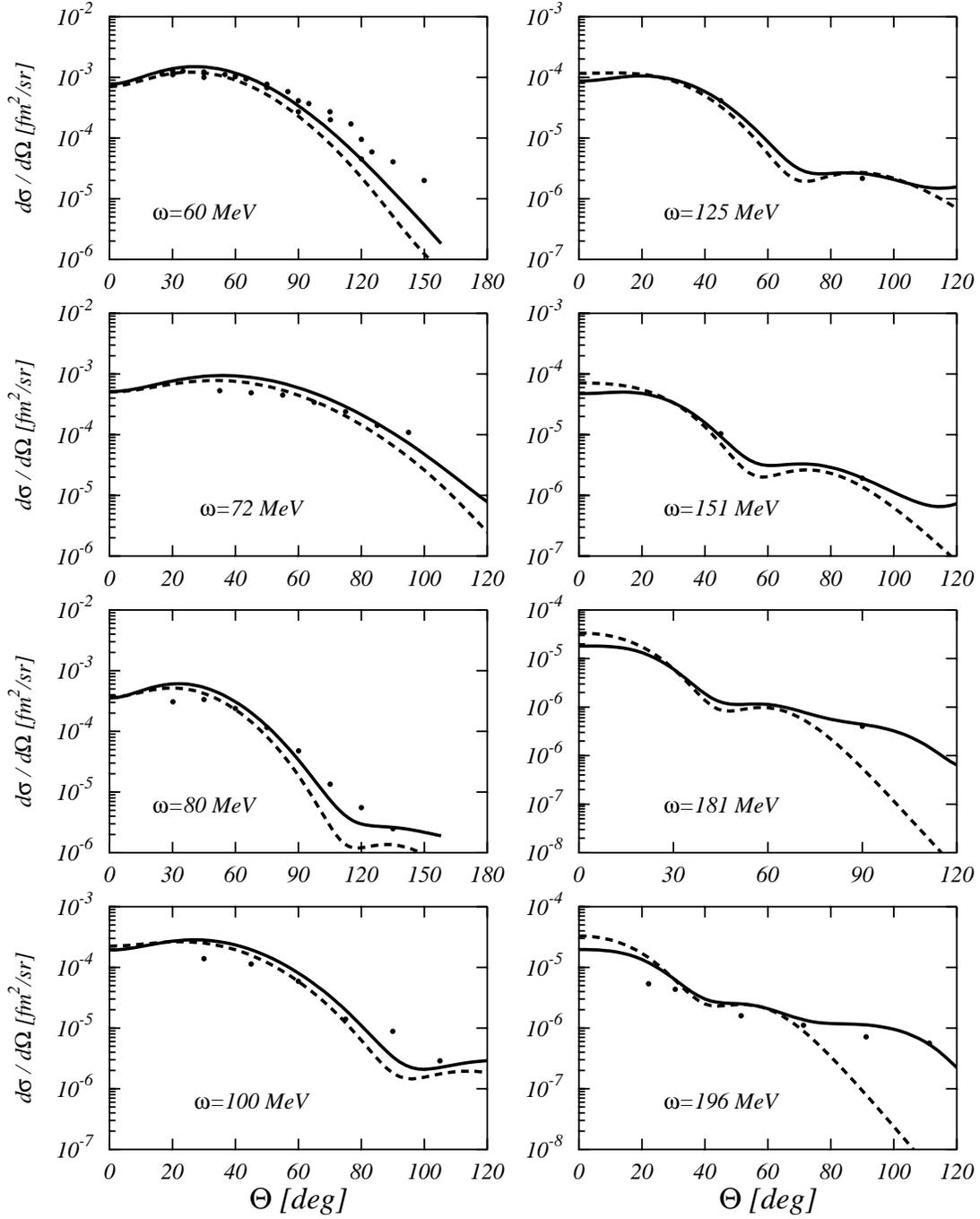}
%\vspace{-2.0cm}
\caption{\small  
  $^{16}$O($\gamma$,p$_0$)$^{15}$N 
  cross sections for
  various values of the photon energies. All the curves have been
  multiplied by the spectroscopic factor of 0.8. The dashed lines show
  the IPM results with OB currents only
  and the full lines the results obtained by including both
  SRC and MEC. Data as in Fig. \protect\ref{fig:gpcor}.
}
\label{fig:gpsf}
\end{figure}
%%%%%%%%%%%%%%%%%%%%%%%%%%%%%%%%%%%%%%%%%%%%%%%%%%%%%%%%%%%%%%%%%%%%%%

In the panel $(b)$ of Fig. \ref{fig:eep} we compare the results of our
calculations, including the spectroscopic factor, with the (e,e'p)
data taken at Mainz \cite{blo95}. The data at low $p_i$ value are
rather well reproduced, as it was expected because of the fit of the
NIKHEF data covering the same range of $p_i$.  From 200 to 300 MeV/c
the theory overestimates the data, but theory and experiment have again
a reasonable agreement at higher values of $p_i$. Also in this case the
need of SRC and MEC to enhance the high momentum tail is evident.

Using the spectroscopic factor value of 0.8 we compare again the
results of our calculations with the ($\gamma$,p$_0$) data in Fig.
\ref{fig:gpsf}. In this figure the dashed lines represent the IPM
results and the full lines the results of the calculations when SRC
and MEC are included.
The agreement with the data has certainly improved with respect to the
results shown in Figs. \ref{fig:gptot} and \ref{fig:gptot1}. 
The inclusion of both MEC and SRC is necessary to obtain the correct
shape of the data.

\section{SUMMARY AND CONCLUSIONS}
\label{conclusion}
We have investigated the photonuclear cross section above the giant
resonance region with the model presented in Ref. \cite{co01}, which
considers the contribution of the SRC at the first order in the
correlation line. The application of our model has been focused on the
$^{16}$O nucleus.

We first studied the importance of the collective nuclear excitations
in the energy region above the giant resonance. This has been done by
calculating the total photoabsorption cross sections with the
Fourier-Bessel continuum RPA approach
of Refs. \cite{co85,deh82}. We have used
two different residual interactions, a zero range Landau--Migdal
interaction and the finite range polarization potential of Ref.
\cite{pin88}. The results obtained did not differ very much, and
showed that, while the position of the giant resonance is rather well
reproduced, its width is too narrow. All our continuum
RPA calculations above the giant resonance region underestimate the
experimental total photo-absorption cross section. The IPM produces
results which are more than one order of magnitude smaller than those of
the RPA. The comparison with the ($\gamma$,p) angular distributions
for photon energies above 60 MeV is very unsatisfactory.

We found that, above the giant resonance region, the ($\gamma$,p)
cross section calculated in the IPM and in RPA are similar while
the ($\gamma$,n) results are quite different.
Since our model cannot describe
collective excitations,
we restricted our investigation to the ($\gamma$,p)
reaction above the giant resonance.
Following the treatment commonly adopted to describe the  (e,e'p)
reactions, we have used an energy dependent
optical potential to treat the FSI.

With this model we have investigated the role of the various terms of the
electromagnetic current. We studied the difference between
electron and photon scattering processes by considering the transverse
response at fixed energy for various values of 
the momentum transfer. We found that at the photon
point ($q=\omega$) the response is strongly dominated by the
convection current, while with increasing $q$ the magnetization
current becomes quickly more important. However,
even at the photon point it is
not possible to neglect the magnetization current for energies
above 100 MeV.

The MEC have been analyzed following the model of Ref. \cite{ama93},
where the pion-exchange diagrams of Fig. \ref{fig:fey}, labelled as
seagull, pionic and $\Delta$, have been  included. 
The use of the seagull diagram only, 
as it is done in Ref. \cite{ben94,gai00}, overestimates 
the MEC effects. The inclusion of the pionic diagram reduces
this effect, and the $\Delta$ current is important only for photon
energies above 100 MeV. 
At these energies, a large difference between
cross sections calculated with OB currents only and those obtained with
the inclusion of the MEC is found for large emission angles. 

Our SRC calculations have been done with three different
correlation functions taken from the FHNC calculations of Refs.
\cite{ari96,fab00}. As already shown for the electron scattering case
\cite{co01,mok01}, the effects of two- and three-point diagrams have
opposite sign.
We observed that the
photoemission cross sections show larger sensitivity to the
correlations than the electron scattering cross sections.
This is due to the different momentum dependences of the transition
matrix elements of the IPM and
of those containing the correlation function.
The IPM transition matrix elements have a sharp
rise, while the other ones are almost flat. At the photon point the
difference between the two transition matrix elements is relatively
smaller than at higher values of the momentum transfer. This is the
source of this relatively high sensitivity.

We found that at excitation energies above 150 MeV, the
cross sections obtained in the IPM plus OB currents framework, 
have a sharp decrease for large values of the nucleon emission
angle. In this region,
the correlations produce cross sections order of
magnitude larger than the IPM ones. This is exactly the same, well known,
behavior of the nucleon momentum distribution. In effect, converting
the emission angle value with the initial momentum of
the emitted nucleon, it is possible to verify that IPM and correlated cross section
start to separate at about 600 MeV/c, where also the momentum
distributions of $^{16}$O start to separate.
Unfortunately, in these cross sections the MEC play an important
role, more important than that of the SRC. Therefore the effect of SRC
is overwhelmed by that of the MEC. 

The comparison with the data improves when a spectroscopic factor is
used. We fixed the value of 0.8 for the 1p$_{1/2}$ proton emission, by
applying our model to the $^{16}$O(e,e'p$_0$)$^{15}$N 
data taken at NIKHEF \cite{leu94}. 
The same spectroscopic factor value has been used 
to describe the coincidence
electron scattering data taken at Mainz \cite{blo95} and the
photon emission data available in the literature. The inclusion of
this reduction factor greatly improves the agreement with all
the data sets, in
spite of the fact that the factor was extracted for relatively low
values of the missing momentum.

We would like to conclude by mentioning some possible way of
improving our model. 
A first one is related to the treatment of the
MEC. The energy values of interest is above the pion emission threshold.
One may
claim that this channel has just open and therefore it is negligible, but
one would like to have an estimate of its relevance. 
A second possible improvement is related to the optical potential at
present taken from literature. In a more consistent description the
potential should be related to the same nuclear structure model
producing the SRC.

\end{document}